\documentclass[apj,numberedappendix]{emulateapj}
\usepackage{mathptmx}
\usepackage{mathrsfs}

\newcommand{\Vec}[1]{\mathbf{#1}}

\journalinfo{The Astronomical Journal}

\begin{document}

\title{Identikit 1: A Modeling Tool for Interacting Disk Galaxies}

\author{Joshua E. Barnes}
\affil{Institute of Astronomy, University of Hawaii, \\
       2680 Woodlawn Drive, Honolulu, HI 96822, USA \\
       \email{barnes@ifa.hawaii.edu}}

\and

\author{John E. Hibbard}
\affil{National Radio Astronomy Observatory, \\
       520 Edgemont Road, Charlottesville, VA 22903, USA \\
       \email{jhibbard@nrao.edu}}

\begin{abstract}
By combining test-particle and self-consistent techniques, we have
developed a method to rapidly explore the parameter space of galactic
encounters.  Our method, implemented in an interactive graphics
program\footnote{This software is available at
\url{http://www.ifa.hawaii.edu/faculty/barnes/ \linebreak
research/identikit/}~.}, can be used to find the parameters
required to reproduce the observed morphology and kinematics of
interacting disk galaxies.  We test this system on an artificial
data-set of $36$ equal-mass merging encounters, and show that it is
usually possible to reproduce the morphology and kinematics of these
encounters and that a good match strongly constrains the encounter
parameters.
\end{abstract}


\section{Introduction}

The diverse morphological and kinematic features of interacting disk
galaxies have a simple dynamical explanation: galactic bridges and
tails \citep{TT72}, rings \citep{LT76, TS77} and related structures
result when ordinary galactic disks experience strong tides in close
encounters.  Tides also cause interacting galaxies to merge by
inexorably transferring energy and momentum from relative motion to
internal degrees of freedom \citep{T77, W78, B88}.  With such a
straightforward physical basis, one might expect that dynamical
modeling of interacting galaxies would be relatively easy.  However,
it's very time-consuming to explore the large parameter space required
to describe a galaxy collision and find a good match to the kinematics
and morphology of a specific system.
In addition, it's never been entirely clear that a good match yields a
unique or physically meaningful model.

In this paper we develop and test an efficient methodology to model
the observable morphology and kinematics of pairs of interacting disk
galaxies.  Empirical tests show that the resulting models can be used
to make strong inferences about the systems they match.  While we do
not consider minor mergers and interactions in this paper, our
methodology can easily be extended to treat such encounters.

At first glance, it seems all too easy to model interacting galaxies
-- and impossible to do so with any degree of confidence.  The
dynamical state of a galactic collision is described by a phase-space
{\it distribution function\/}, $f(\Vec{r},\Vec{v})$, which gives the
mass density at position~$\Vec{r}$ and velocity~$\Vec{v}$.  In
contrast, observations of a specific component $c$ (e.g., neutral
hydrogen) yield a {\it data cube\/} $F_c(X,Y,V)$, which represents the
distribution of that component at each point $(X,Y)$ on the plane of
the sky as a function of line-of-sight velocity $V$.  Since
$f(\Vec{r},\Vec{v})$ depends on {\it six\/} variables, while
$F_c(X,Y,V)$ depends on only {\it three\/}, it appears that
observations do not provide enough information.  Put simply, there are
an infinite number of different 6-D distribution functions consistent
with any given 3-D data cube.

On further reflection the problem is not quite as hopeless as it
appears.  This is because a typical galaxy merger begin with a tidal
encounter between two normal, fairly symmetric spirals.  Galaxies are
scrambled as they merge, but the stars and dark matter which
constitute most of their mass evolve collisionlessly.  The fundamental
dynamical equation,
\begin{equation}
{\partial f \over \partial t} +
  \Vec{v} \cdot {\partial f \over \partial \Vec{r}} -
    {\partial \Phi \over \partial \Vec{r}} \cdot
      {\partial f \over \partial \Vec{v}} = 0, 
\end{equation}
where $\Phi$ is the gravitational potential, is fully reversible
\citep[e.g.][]{vAvG77}; thus in some sense the original galaxies are
still there, imposing a hidden symmetry on the dynamical state of a
merging system.  So we need not consider {\it all\/} possible
distribution functions consistent with a given data cube; only a very
small subset of these functions can possibly result from an encounter
between two normal disk galaxies.

In practice, mergers are modeled by guessing initial conditions,
numerically simulating the ensuing collision, and comparing the result
to the morphology and kinematics of the system one wants to model.  If
the model fails to match the observations, go back and guess again
until the results are satisfactory.  Most of the guess-work focuses on
selecting the disk orientations, typically specified by angles
$(i_1,\omega_1)$ and $(i_2,\omega_2)$ for disks 1 and 2 \citep{TT72};
also needed are the eccentricity $e$ and pericentric separation $p$ of
the initial orbit, as well as the galactic mass ratio $\mu$.  So {\it
seven\/} parameters are needed to specify the initial conditions for
an encounter of two axisymmetric disk galaxies -- not counting the
parameters used to specify their internal
structures\footnote{Selecting the correct internal structures is a
separate problem, and one largely beyond the scope of this paper.
Under some fairly general assumptions, the internal structure of an
axisymmetric galaxy may be described by a distribution function $f =
f(E, J_\mathrm{z}, I_3)$ depending on the energy $E$, angular momentum
about the symmetry axis $J_\mathrm{z}$, and a third integral of motion
$I_3$.  Formally speaking, an {\it infinite\/} number of parameters
are needed to specify such a function.}.

Once a simulation has been run, one must select another {\it nine\/}
parameters when comparing the results to observational data: a time
$t$ since pericenter, a viewing direction given, for example, by
angles $(\theta_{\rm X},\theta_{\rm Y},\theta_{\rm Z})$, scale factors
$\mathscr{L}$ and $\mathcal{V}$ for length and velocity\footnote{If
the simulation is conducted in physical units then these parameters
are not necessary -- but additional parameters are required to
describe the initial conditions, so the total parameter count is
unchanged.}, and a center-of-mass position on the plane of the sky
$(X_{\rm cm},Y_{\rm cm})$ and velocity $V_{\rm cm}$.
All told, a minimum of {\it sixteen\/} parameters are needed to
completely specify the initial conditions, time, point of view, and
scale of a merger model.  This plethora of parameters has long posed a
challenge for systematic surveys of galactic collisions \citep{TT72,
FS82, WS92, HKBB93, B98, NB03}.  The problem we address here is
slightly different -- instead of trying to survey the entire parameter
space, we want to navigate toward a solution matching the morphology
and kinematics of a given interacting system.  Intuition and prior
experience can guide this process by narrowing the range of parameter
space explored.  Nonetheless, given the size and complexity of this
parameter space, it's not surprising that many simulations must be run
to attempt a match, or that detailed models of galactic collisions are
not easy to produce.

\section{``Identikit'' Methodology}
\label{sec:methodology}

While self-consistent simulations are useful to finalize dynamical
models of tidally interacting galaxies, here we simulate galactic
disks with test particles.
Test particles have a long history \citep{PS61,TT72,CFWG80,HQ87,WS92}
and nicely reproduce features such as bridges, tails, and shells which
develop with little direct influence from self-gravity.
To include orbital decay, which is crucial in modeling the more
advanced stages of galaxy encounters and mergers, the test particles
may be used to estimate the drag on the central masses
\citep[e.g.,][]{TT72, B84, QG86}.  However, orbit decay is largely
driven by tidal interactions of galaxy {\it halos\/}
\citep[e.g.,][]{T77, W78, B88}, and it's relatively easy to compute
the self-consistent interaction of two halos using N-body simulations.
Our initial approach was therefore to represent the mass of each
galaxy with a spherical distribution of massive particles; in each of
these spheres, we embedded {\it multiple\/} disks of test particles,
and decided which disk to display {\it after\/} running the
simulations\footnote{In forensic investigations, ``Identikit'' is one
of several systems used to construct portraits by selecting from a
menu of facial features.  Our approach is analogous.}.

Building on this idea, and taking advantage of the faster processors
now available, we have replaced these discrete collections of disks
with spherical swarms of test particles moving on circular orbits --
in effect, populating each galaxy model with {\it all\/} possible
disks.

The modeling procedure begins by selecting mass models for the two
galaxies, thereby fixing the mass ratio $\mu$.  Our models include a
bulge, a disk, and a halo; these components have cumulative mass
profiles $m_{\rm b}(r)$, $m_{\rm d}(r)$, and $m_{\rm h}(r)$,
respectively, where $m_c(r)$ is the total mass in component $c$ within
radius $r$.  For each galaxy, we compute the total mass profile
\begin{equation}
  m(r) = m_{\rm b}(r) + m_{\rm d}(r) + m_{\rm h}(r) \, ,
  \label{eq:total-mass-prof}
\end{equation}
calculate the corresponding isotropic distribution function using
Eddington's (\citeyear{E16}) formula \citep[e.g.,][p.~236]{BT87}, and
construct a spherical N-body realization of this profile using $N_{\rm
sphr}$ equal-mass particles.  Details of this procedure are given in
Appendix~A.

Each of these spherical N-body realizations is then loaded with
$N_{\rm test}$ test particles moving on circular orbits.  We pick the
orientation of the orbit of test particle $i$ by randomly drawing its
normalized angular momentum $\hat{\Vec{s}}_i$ from a uniform
distribution on the unit sphere $\mathbf{S}^2$.  The radial
distribution of the test particles may be chosen at will.  The
simplest choice is to use a radial distribution following the
cumulative profile of the disk, $m_{\rm d}(r)$; with this approach,
the test particles representing a disk with a normalized spin vector
$\hat{\Vec{s}}_{\rm d}$ are those with
\begin{equation}
  1 - \hat{\Vec{s}}_{\rm d} \cdot \hat{\Vec{s}}_i \le \sigma \, ,
  \label{eq:select_unbiased_disk}
\end{equation}
where $\sigma \ll 1$ is a tolerance parameter proportional to the
number of particles selected.  In practice, however, this places many
disk particles at small radii where they are largely immune to tides.
The sampling at large radii can be improved by radially biasing the
test particle distribution.  We do this by multiplying the local disk
particle density, $\rho_{\rm d}(r) = (4 \pi r^2)^{-1} d m_{\rm d}/d
r$, by a factor of $r^2$, and replacing
(\ref{eq:select_unbiased_disk}) with
\begin{equation}
  1 - \hat{\Vec{s}}_{\rm d} \cdot \hat{\Vec{s}}_i \le
    \sigma \, / \, \mathrm{max}(q_i, r_{\rm min})^2 \, .
  \label{eq:select_biased_disk}
\end{equation}
Here $q_i$ is the {\it initial\/} orbital radius of particle $i$, and
$r_{\rm min}$ is a parameter which keeps (\ref{eq:select_biased_disk})
from diverging for small $q_i$.  Particles selected using
(\ref{eq:select_biased_disk}) follow the original disk distribution
down to radius $r_{\rm min}$, at smaller radii the disk is
undersampled, but this has little effect if $r_{\rm min}$ is small.

Finally, two such configurations are placed on a relative orbit with a
given pericentric separation $p$ and eccentricity $e$ and followed
until they merge; we save particle positions and velocities every few
time-steps, creating a data-base of several hundred frames tracing the
system's history from start to finish.  This data can then be used to
approximate {\it any\/} encounter with the chosen $\mu$, $p$, and $e$.

\begin{figure*}[t!]
\begin{center}
\includegraphics[clip=true,width=0.85\textwidth]{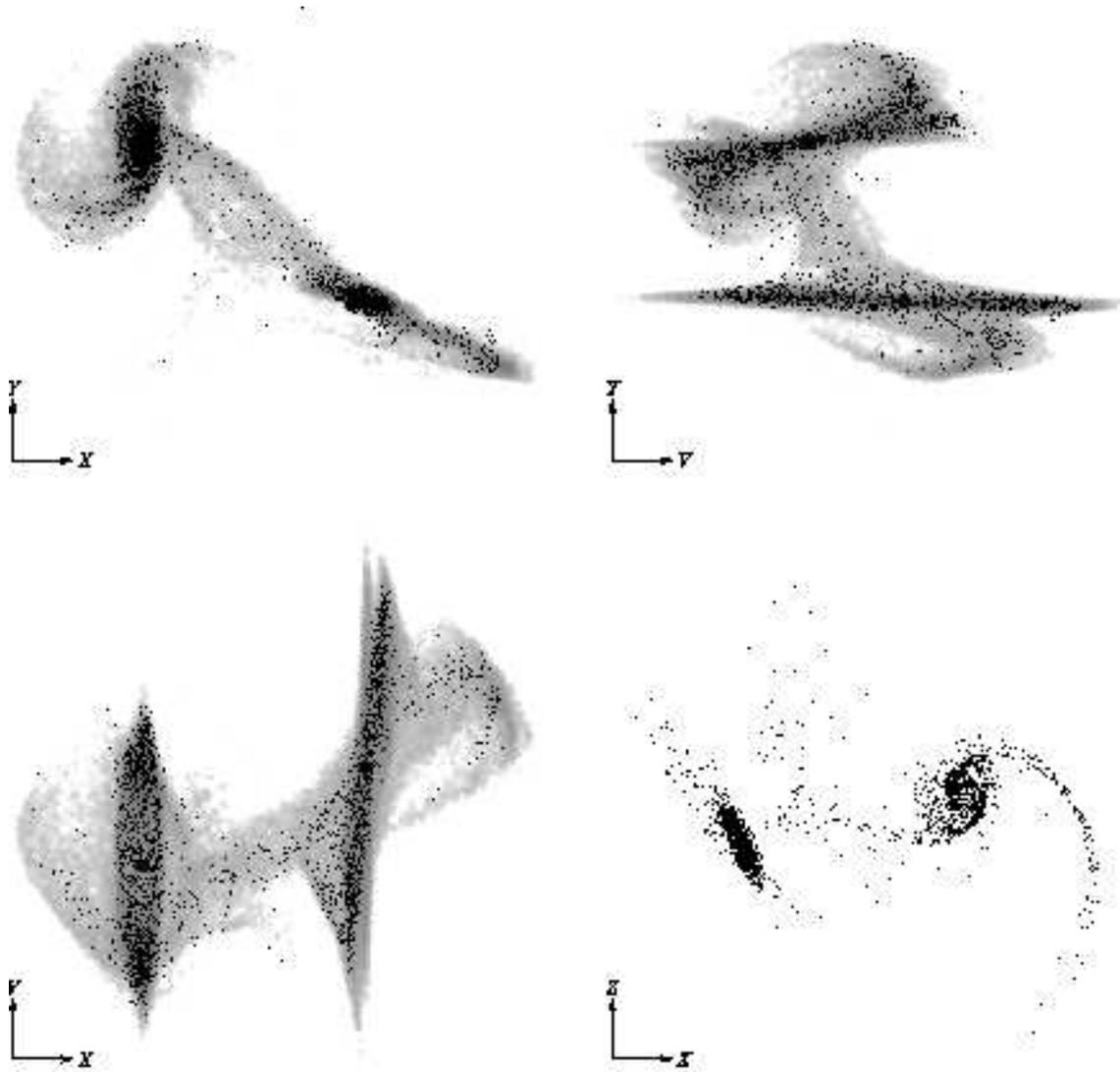}
\caption{Identikit match to a pair of merging galaxies.  Grey-scale
  images represent the data cube $F(X,Y,V)$ of the system, while
  points represent the Identikit model.  Top-left, top-right, and
  bottom-left quadrants show the data cube and the model projected on
  the $(X,Y)$, $(V,Y)$, and $(X,V)$ planes, respectively; bottom-right
  quadrant shows the model projected on the $(X,Z)$ plane.
  \label{idkit_match}}
\end{center}
\end{figure*}

Identikit software includes interactive routines allowing the user to
select the disk orientations, viewing direction, scale factors, and
centroid positions; the resulting test-particle coordinates are
instantly projected on the $(X,Y)$, $(X,V)$, $(V,Y)$, and $(X,Z)$
planes.  The user can also step forward or backward in time, switch
between data-bases created using different values for $p$, $\mu$, or
$e$, and vary the tolerance parameter $\sigma$.  In addition, an
observational data cube $F(X,Y,V)$ for a specific system to be matched
may be projected onto the $(X,Y)$, $(X,V)$, and $(V,Y)$ planes,
typically using grey-scale or contour images.  A lap-top computer can
easily store the necessary data and supply the modest processing power
required when searching for a match.  Fig.~\ref{idkit_match} presents
an example; the encounter and viewing parameters have been adjusted so
the test particles (points) closely match the data cube (grey-scale).

\section{Experimental Validation}

The Identikit system is designed with the primary goal of analyzing
observations.  However, it's not straightforward to perform empirical
tests using observational data.  For one thing, much of the available
data is rather heterogeneous; data-sets with uniformly high resolution
and signal-to-noise ratio are not easy to obtain.  For another, we
don't know the true initial conditions and viewing parameters for more
than a handful of the mergers that have been observed, so we don't
have any simple way to validate the results of our modeling.  To
determine if Identikit can actually reconstruct galactic encounters
from the information contained in data-cubes, we tested it on an
artificial data set of $36$ self-consistent disk galaxy merger
simulations with random orientations, times since first passage,
viewing directions, and scale factors.

\subsection{Artificial Merger Data}

Our disk galaxy model has a spherical bulge \citep{H90a} containing
$5$\% of the mass, an exponential/isothermal disk \citep{dV59a, dV59b,
F70, vdKS81} containing another $15$\%, and a spherical dark halo
\citep{NFW96} containing the remaining $80$\%.  The density profiles
for these components are
\begin{equation}
  \begin{array}{@{}l}
    \rho_{\rm b}(r) \propto
      r^{-1} (r + a_{\rm b})^{-3} \,, \\
    \rho_{\rm d}(q,z) \propto
      e^{- q / a_{\rm d}} \, \mathrm{sech}^2(z/z_{\rm d}) \,, \\
    \rho_{\rm h}(r) \propto
      r^{-1} (r + a_{\rm h})^{-2} \,,
  \end{array}
\end{equation}
where $a_{\rm b}$ is the scale length of the bulge, $q = \sqrt{x^2 +
y^2}$ is the cylindrical radius, $a_{\rm d}$ is the scale length of
the disk, $z_{\rm d}$ is the scale height of the disk, and $a_{\rm h}$
is the scale length of the halo.  Each galaxy was realized using a
total of $N = 131072$ particles.  The simulations used natural units
with Newton's constant $G = 1$.  In these units, the galaxy model has
total mass $m = 1.25$ and half-mass radius $r_{\rm med} \simeq 0.532$.
The disk's scale length $a_{\rm d} = 1/12$, and the median circular
velocity of the disk material is $v_{\rm med} \simeq 1.23$; at a
radius of $3 a_{\rm d}$ the orbital period is $t_{\rm orb} \simeq
1.23$.

We restricted our artificial data set to equal-mass ($\mu = 1$)
encounters with parabolic initial orbits ($e = 1$); for each orbit,
the pericentric separation $p$ was drawn from a uniform distribution
in the range $[0.05,0.5] = [0.6,6] a_{\rm d}$.  The instant when this
idealized two-body orbit reaches pericenter defines $t = 0$; times $t
< 0$ are before pericenter, while times $t > 0$ are after pericenter.
We adopt a coordinate system in which the orbital angular momentum
vector is parallel to the $\hat{\Vec{z}}$ axis.  The normalized spin
vector $\hat{\Vec{s}}$ of each disk was chosen from a uniform
distribution on the unit sphere $\mathbf{S}^2$; in practice, the
inclination $i$ was chosen by drawing $\cos(i) = \hat{\Vec{s}} \cdot
\hat{\Vec{z}}$ from a uniform distribution in the range $[-1,1]$, and
the argument $\omega$ was chosen from a uniform distribution in the
range $[0^\circ,360^\circ]$.  Further details on the galaxy models and
the merger simulations are given in Appendix~B.

For each of the $36$ merger simulations we chose a random time between
first and second pericenter, rescaled the system by random factors in
length and velocity, and ``observed'' it from a random direction.  We
first determined the relative orbit of each pair of galaxies, using
the most tightly-bound $2048$ particles in each bulge to measure
galactic positions.  Let $t_1 \simeq 0$ and $t_2$ be times of first
and second pericenter, respectively; the random time $t$ was drawn
from a uniform distribution in the range $[t_1, t_2]$ and rounded down
to the nearest available output time.

Next, we selected scale factors $\mathscr{L}$ and $\mathcal{V}$ for
length and velocity, respectively.  These were chosen so that the
galaxy models obey a mass-radius-velocity relation of the form $M
\propto R^2 \propto V^4$ with a small amount of scatter
\citep[c.f.][]{TF77}.  Let $\xi$ be drawn from a uniform distribution
in the range $[-0.5,0.5]$, and $g_1$ and $g_2$ be drawn from a
Gaussian distribution with zero mean and unit dispersion; then
\begin{equation}
    \mathscr{L} = 10^{\xi/2} \, 10^{0.05 g_1} \, ,
    \qquad
    \mathcal{V} = 10^{\xi/4} \, 10^{0.05 g_2} \, .
\end{equation}

Finally, we chose a random viewing direction $\widehat{\Vec{Z}}$ from
a uniform distribution on the unit sphere $\mathbf{S}^2$.  We drew a
second vector $\widehat{\Vec{X}}_0$ from the same distribution, and
set $\widehat{\Vec{X}} = \widehat{\Vec{X}}_0 - \widehat{\Vec{Z}}
(\widehat{\Vec{X}}_0 \cdot \widehat{\Vec{Z}})$ and $\widehat{\Vec{Y}}
= \widehat{\Vec{Z}} \times \widehat{\Vec{X}}$.  These vectors and
scale factors were used to map the position $\Vec{r}_i$ and velocity
$\Vec{v}_i$ of each particle $i$ to data-cube coordinates:
\begin{equation}
    X_i = \mathscr{L} \, \widehat{\Vec{X}} \cdot \Vec{r}_i \, , \qquad
    Y_i = \mathscr{L} \, \widehat{\Vec{Y}} \cdot \Vec{r}_i \, , \qquad
    V_i = \mathcal{V} \, \widehat{\Vec{Z}} \cdot \Vec{v}_i \, .
\end{equation}

Particles from the disks of the two galaxies, transformed to $(X_i,
Y_i, V_i)$ coordinates, provide an N-body representation of a data
cube $F_{\rm d}(X,Y,V)$ for the disk material.  Such data is roughly
comparable to the neutral hydrogen data-cubes $F_{\rm HI}(\alpha,
\delta, V_{\rm los})$ available for many interacting galaxies
\citep[e.g.,][]{HvGRS01}.  The simulated data has better resolution
than most observational data-sets and is free of noise and
interferometric artifacts; moreover, our simulations used
collisionless particles instead of neutral gas.  We could have run
random mergers with gas to improve the correspondence between the
simulations and real observational data, but the computing time
required for a large suite of simulations with gas is non-trivial.
Fortunately, collisionless simulations do a good job of reproducing
the tidal features commonly detected in H{\small I} since the latter
usually evolve ballistically once tidally extracted from their parent
galaxies.

To present the simulated data in the form required for Identikit
matching, we projected the disk particles for each of our $36$ mergers
on the $(X,Y)$, $(X,V)$, and $(V,Y)$ planes; gridded particle
distributions were lightly smoothed to produce grey-scale images.  We
also used tightly-bound particles from the bulge of each galaxy to
determine its position and line-of-sight velocity; the results were
plotted on top of the grey-scale images.  Our images are thus
analogous to H{\small I} maps supplemented with accurate nuclear
coordinates and velocities.

\begin{figure*}[p!]
\begin{center}
\includegraphics[clip=true,width=0.95\textwidth]{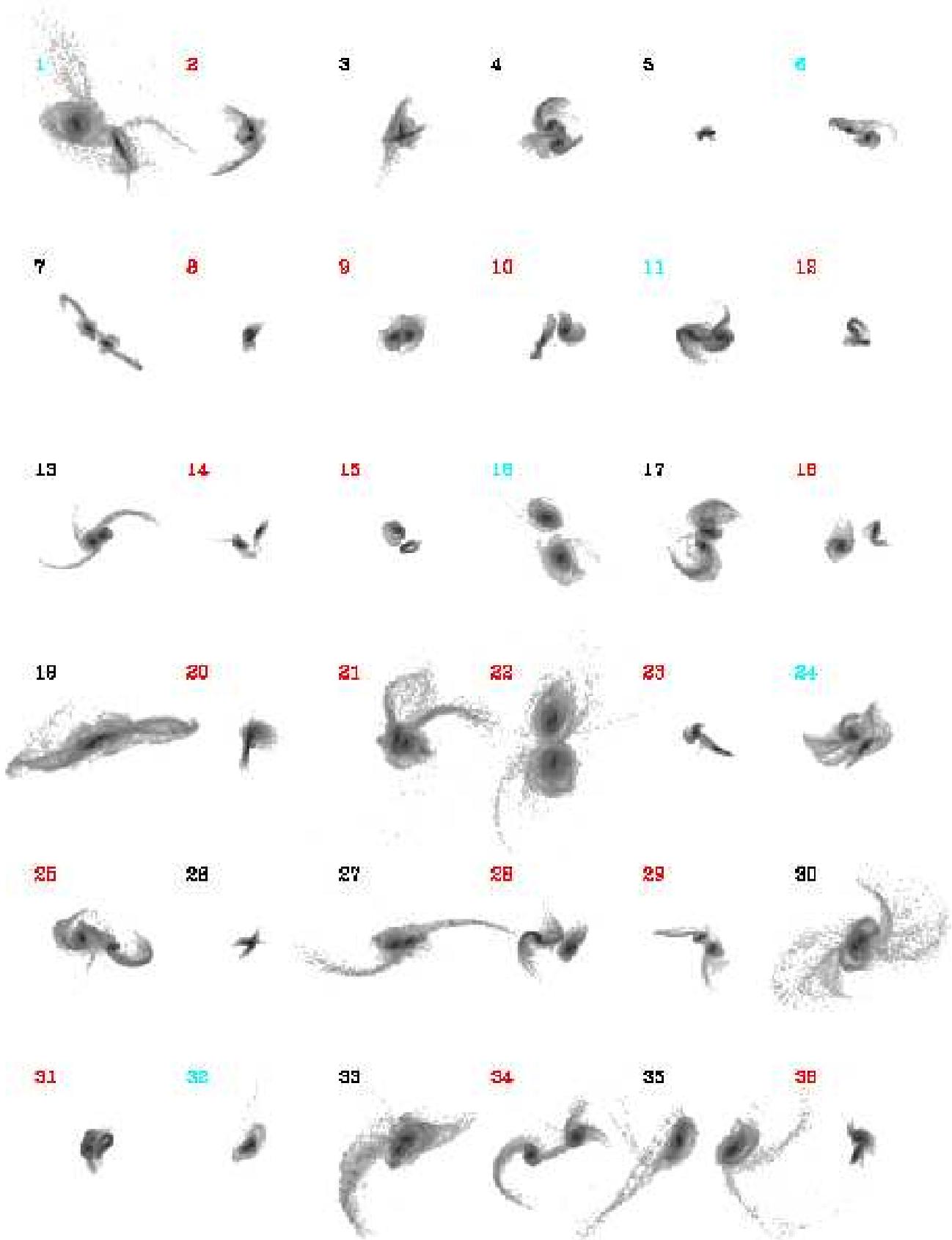}
\caption{Sky-plane $(X,Y)$ projections of the $36$ merging encounters
  used to test the Identikit procedure.  Note the range of sizes and
  morphologies.  The number of each system appears to its upper left;
  color indicates Identikit fit quality (see \S~\ref{sec:matching}),
  with good fits in red, fair fits in black, and poor fits in blue.
  \label{random_mergers}}
\end{center}
\end{figure*}

The entire procedure outlined above, including both the generation of
the simulations and the selection of viewing parameters, was performed
by automated scripts without human intervention; we did not know the
actual values of any parameters except $e$ and $\mu$.  The resulting
sample of merging galaxies, shown in Fig.~\ref{random_mergers},
possess a variety of morphologies; only a subset display the ``double
tails'' characteristic of the best-known mergers \citep{T77}.

\subsection{Identikit Matching}
\label{sec:matching}

We prepared a series of eight Identikit simulations spanning a range
of pericentric separations.  Each simulation contained two identical
($\mu = 1$) configurations of $N_{\rm test} = 262144$ test particles
and $N_{\rm sphr} = 81920$ massive particles; the mass model used to
set up the massive particle distribution was a spherical version of
the one used in the random mergers.  These configurations were placed
on parabolic ($e = 1$) relative orbits with pericentric separations $p
= 1/16$, $2/16$, $3/16$, $4/16$, $5/16$, $6/16$, $7/16$, and $8/16$,
starting at $t = -2$ time units before first pericenter, and followed
until $t = 8$, by which time even the widest passage had merged.

We used these simulations and the Identikit software to fit each of
the random mergers by interactively matching the ``observed'' $(X,Y)$,
$(X,V)$, and $(V,Y)$ projections with test particles. The modeling
process usually began with rough guesses for the viewing direction,
time since pericenter, and pericentric separation.  A variety of clues
guided these guesses.  For example, short but pronounced tidal
features point to a recent tidal encounter, while long but attenuated
features suggest an older passage; in later stages, loops associated
with tails show that material has started falling back.  Likewise, if
the galaxies display a large separation in projected velocity then the
sight-line must be close to the orbital plane; conversely, a small
difference in projected velocity implies either that the system is
observed near apocenter or that the relative velocity vector is
roughly perpendicular to the line of sight.  Finally, other things
being equal, closer passages generally yield stronger and more
dramatic tidal features.

The next step was to adjust the orientations of the two disks,
attempting to roughly match the morphology and kinematics of the
system. This generally suggested further modifications to the viewing
direction, separation, and time, as well as the scale factors and
center-of-mass position and velocity.  Further adjustment of all
parameters continued until a satisfactory match was obtained or
exhaustion set in.  Our criteria for a satisfactory match were
somewhat subjective\footnote{It's not trivial to evaluate matches
quantitatively; for more on this, see \S~\ref{sec:genetic}.}; we
placed a good deal of weight on matching tidal features (e.g.,
Fig.~\ref{idkit_match}), while recognizing that test particles can't
accurately reproduce structures -- such as tidally-induced spirals --
which depend on self-gravity.
Each match typically took a few hours, and the entire set of $36$
random mergers was matched in about one month; for comparison, a match
to the NGC~7252 merger remnant \citep{HM95} took $74$ N-body runs over
a three month time period, while a match to the NGC~4676 system
\citep{B04} took $\sim 30$ runs over two months.

After changing the viewing direction, time since pericenter or orbital
parameters, it's usually necessary to reposition the centers of the
models on top of the actual positions by adjusting the rotation about
the viewing axis $\theta_{\rm Z}$, scale factor $\mathscr{L}$, and
center of mass position $(X_{\rm cm}, Y_{\rm cm})$.  We therefore
implemented an option to ``lock'' the centers; when this option is
invoked, $\theta_{\rm Z}$, $\mathscr{L}$, and $(X_{\rm cm}, Y_{\rm
cm})$ are recalculated on the fly, keeping the projected positions of
the models invariant as other parameters are changed.  Locking works
quite well when the two galaxies are well-separated on the $(X,Y)$
plane; it's less useful, and can be downright counter-productive, when
the centers appear close together.  In fine-tuning a nearly final
match we sometimes found it useful to unlock the centers, trading off
slight misalignments in central positions for improved matches to
tidal features.

Fig.~\ref{idkit_match}, which shows our match to object~23 (see
Fig.~\ref{random_mergers}), illustrates many aspects of the matching
process.  From the start, it seemed likely that the viewing direction
would be fairly close to the orbital plane, since the two galaxies
have rather different systemic velocities.  The galaxy on the lower
right of the $(X,Y)$ projection appears nearly edge-on, as indicated
by its morphology {\it and\/} its rather large velocity range.  Since
its tidal features lie more or less in the same plane as the disk
itself, it seemed plausible that this galaxy has a relatively small
inclination $i_1$ to the orbital plane, while its companion clearly
has a higher inclination $i_2$ and appears more face-on from our
viewpoint.  The dual-valued velocities along the tail of the edge-on
disk, which produce the ``hook''-shaped feature seen in the $(X,V)$
and $(V,Y)$ projections, suggested that this tail is actually quite
extended, but viewed so as to double back on itself.

For an initial match to this system, we tried a ``middle-of-the-road''
pericentric separation ($p = 0.25$); at a relatively early time ($t =
0.56$) we could roughly match the velocity difference and some aspects
of the morphology and kinematics, including the spiral morphology of
the face-on disk and the hooked tail in the $(V,Y)$ projection.
However, other features of this initial match were less satisfactory.
In the $(X,V)$ projection, the tail doubled back too soon, while in
the $(V,Y)$ projection, the bridge did not span the velocity range
between the galaxies, falling to the left of its ideal position.
Moreover, the more face-on disk, while nicely rendered in the $(X,Y)$
projection, populated regions of phase space which the $(V,Y)$
projection showed to be empty.  Trial and error revealed that wider
passages and later times could repair most of these defects; the
solution shown in Fig.~\ref{idkit_match} uses a $p = 0.5$ passage
viewed at $t = 1$.  The tail in this match, while more extended than
it was initially, is still a bit too short.  Times $t > 1$ yield
longer tails, but the velocity difference between the galaxies becomes
too small, and bridge particles falling through the more face-on disk
over-populate a relatively sparse region of phase-space.  The adopted
solution is therefore a compromise between several competing factors.

After comparing our Identikit models to the morphology and kinematics
of all $36$ random mergers, we subjectively graded the solutions as
``good'' ($18$ cases), ``fair'' ($12$ cases), or ``poor'' ($6$ cases);
these grades are indicated in Fig.~\ref{random_mergers}.  Good
matches, like the one in Fig.~\ref{idkit_match}, strongly constrain
the parameters.  Fair matches generally appear plausible but allow
more latitude in selecting parameter values; this group included
several systems with twin edge-on tidal tails.  Poor matches could be
divided into two groups: systems with weak and diffuse tidal features,
typically resulting from very wide encounters involving retrograde or
highly inclined galaxies (objects~16 and 32), and systems with
pronounced but confusing tidal features (objects~1, 6, 11, and
24).

\subsection{Results: Parameters}
\label{sec:parameters}

\begin{figure}[t!]
\begin{center}
\includegraphics[clip=true,width=0.35\columnwidth]{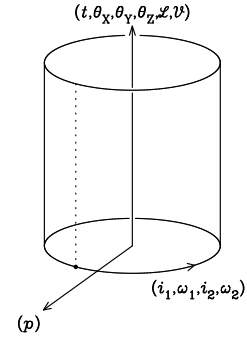}
\caption{An abstract representation of the parameter space of galaxy
  interactions.  The radial coordinate represents the initial orbit,
  the azimuthal coordinate represents the disk orientations, and the
  vertical coordinate represents the parameters chosen after a
  simulation is run. A conventional N-body simulation explores the
  parameter subspace represented by the dotted line, while a single
  Identikit simulation can explore entire cylindrical surface.
  \label{interparam}}
\end{center}
\end{figure}

With the Identikit solutions in hand, we compared their parameter
values to the true values used to generate the artificial merger data.
Fig.~\ref{interparam} uses an abstract cylindrical coordinate system
to portray the parameter space explored in these solutions.  The
radial coordinate represents the initial orbit of the two galaxies;
the fits discussed here parameterize the orbit by the pericentric
separation $p$, since the eccentricity $e$ and mass ratio $\mu$ were
fixed beforehand.  The azimuthal coordinate represents the four angles
$(i_1, \omega_1)$ and $(i_2, \omega_2)$ required to specify the
initial orientations of the two disks.  Together, the radial and
azimuthal coordinates of this abstract space completely specify the
initial conditions for a galaxy interaction.  The vertical coordinate
represents the parameters selected {\it after\/} running a simulation:
time since pericenter $t$, viewing angles $(\theta_{\rm X},\theta_{\rm
Y},\theta_{\rm Z})$, and scale factors $(\mathscr{L},
\mathcal{V})$; here the center-of-mass parameters are omitted since
their values are not discussed below.  A conventional N-body
simulation starts at a point on the horizontal plane and explores the
parameter subspace represented by the dotted line in this figure; a
single Identikit simulation, in contrast, allows access to an entire
cylindrical surface.

\begin{figure*}[t!]
  \begin{minipage}[t]{\columnwidth}
    \includegraphics[clip=true,width=0.94\textwidth]{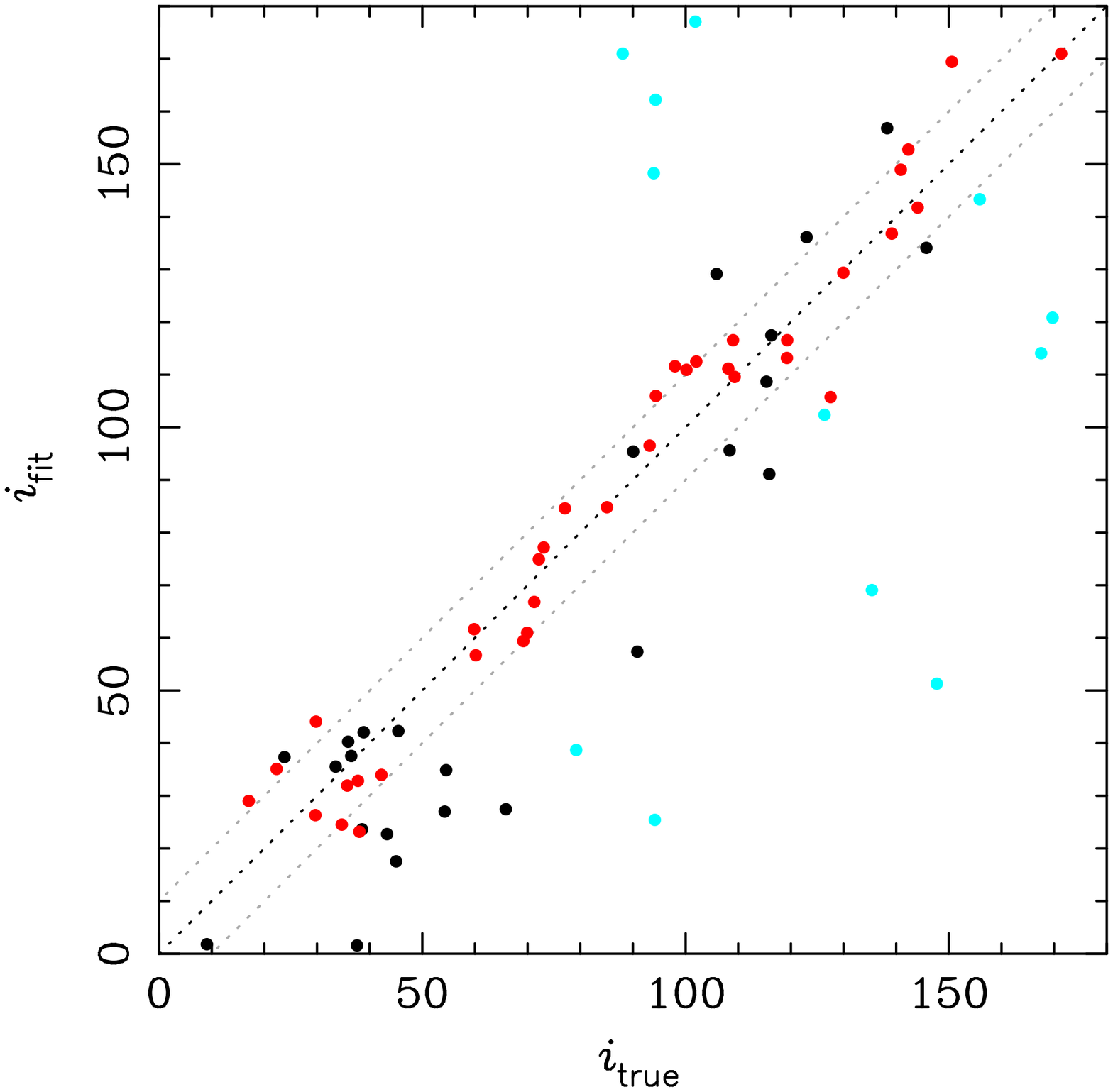}
    \caption{Estimates of disk inclination, $i$.
      The color of each data point shows quality of fit: red
      is good, black is fair, and blue is poor.  The heavy dotted line
      represents perfect agreement ($i_{\rm fit} = i_{\rm true}$); the
      light lines show $i_{\rm fit} = i_{\rm true} \pm 10^\circ$.
      \label{cmp_incl}}
  \end{minipage}%
  \hbox to 0.042\textwidth{}%
  \begin{minipage}[t]{\columnwidth}
    \includegraphics[clip=true,width=0.94\textwidth]{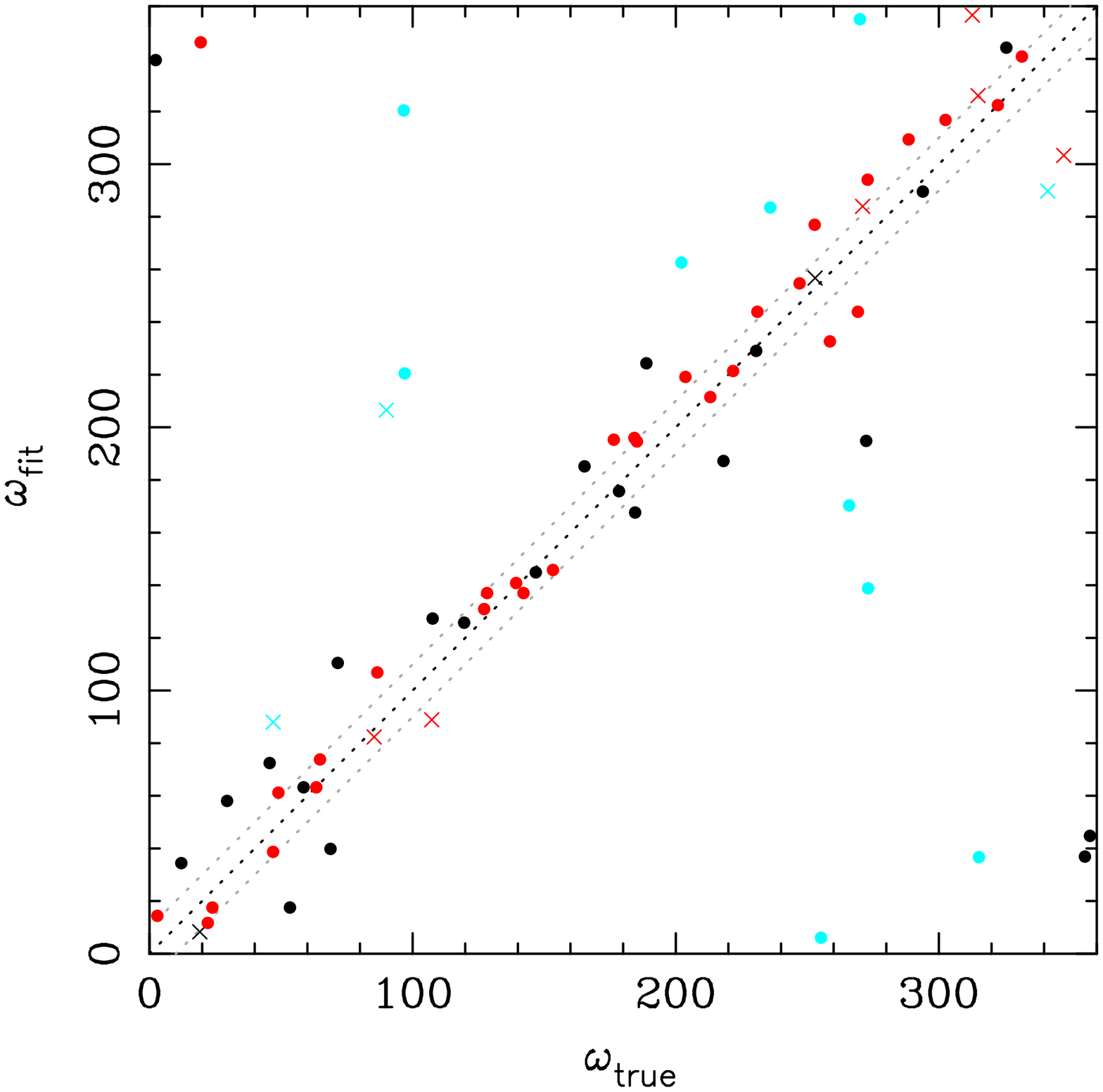}
    \caption{Estimates of disk argument, $\omega$.
      Colors and dotted lines as in Fig.~\ref{cmp_incl}; crosses
      represent disks with $|i_{\rm true} - 90^\circ| > 60^\circ$.
      Note that opposing edges of this plot should be identified.
      \label{cmp_omega}}
  \end{minipage}%
\end{figure*}

Fig.~\ref{cmp_incl} presents a scatter plot comparing inclinations
$i_{\rm fit}$ derived from the Identikit fits against the inclinations
$i_{\rm true}$ used in the random merger sample.  Here and in
subsequent plots, color indicates the grade of each model; note that
the disks are not graded individually, so both disks in a given model
receive the same grade even if one fits better than the other.  The
good fits (shown in red) fall quite close to the diagonal line across
the entire range of the plot.  The fair fits (black) display more
scatter but track the same relationship.  In contrast, the poor fits
(blue) have a very different distribution: almost all have
inclinations $i_{\rm true} > 90^\circ$, and most fall quite far from
the diagonal.  This plot supports a couple of useful inferences.
First, our subjective grades, based on the overall appearance of the
Identikit models, correlate with $|i_{\rm fit} - i_{\rm true}|$; in
other words, these grades mean something.  Second, encounters
involving disks with inclinations $i > 90^\circ$ are more difficult to
model, presumably because the tidal features such encounters produce
are less distinct and more ambiguous.  Nonetheless, many
high-inclination encounters were successfully modeled; other factors
evidently influence the outcome of the modeling process.

Likewise, Fig.~\ref{cmp_omega} compares fit and true values of the
argument to pericenter, $\omega_{\rm fit}$ and $\omega_{\rm true}$.
Since $\omega$ becomes indeterminate for inclinations near $i =
0^\circ$ or $i = 180^\circ$, we plot disks with $30^\circ \le i_{\rm
true} \le 150^\circ$ as filled circles, and disks outside this range
as crosses.  This plot shows that good fits yield arguments quite
close to the true values, and the fair fits do nearly as well.  At
first sight it may seem that a few disks, represented by the one good
and three fair points in the upper left and lower right of the plot,
yield discrepant values of $\omega_{\rm fit}$, but this is an artifact
of topology; $\omega$ is a periodic coordinate, and when opposing
edges of the plot are identified, these apparent outliers are not so
far from $\omega_{\rm fit} = \omega_{\rm true}$.  The poor fits, in
contrast, genuinely scatter throughout the plot.  Most of the good and
fair fits with $i_{\rm true} < 30^\circ$ or $150^\circ < i_{\rm true}$
still yield reasonable values for $\omega$, but these are a bit more
scattered.  For example, only one of the six good fits with $i_{\rm
true}$ in this range yields a $\omega_{\rm fit}$ within $10^\circ$ of
$\omega_{\rm true}$; in contrast, just over half of the good fits with
inclinations $30^\circ \le i_{\rm true} \le 150^\circ$ have
$|\omega_{\rm fit} - \omega_{\rm true}| < 10^\circ$.

\begin{figure*}[t!]
  \begin{minipage}[t]{\columnwidth}
    \includegraphics[clip=true,width=0.94\textwidth]{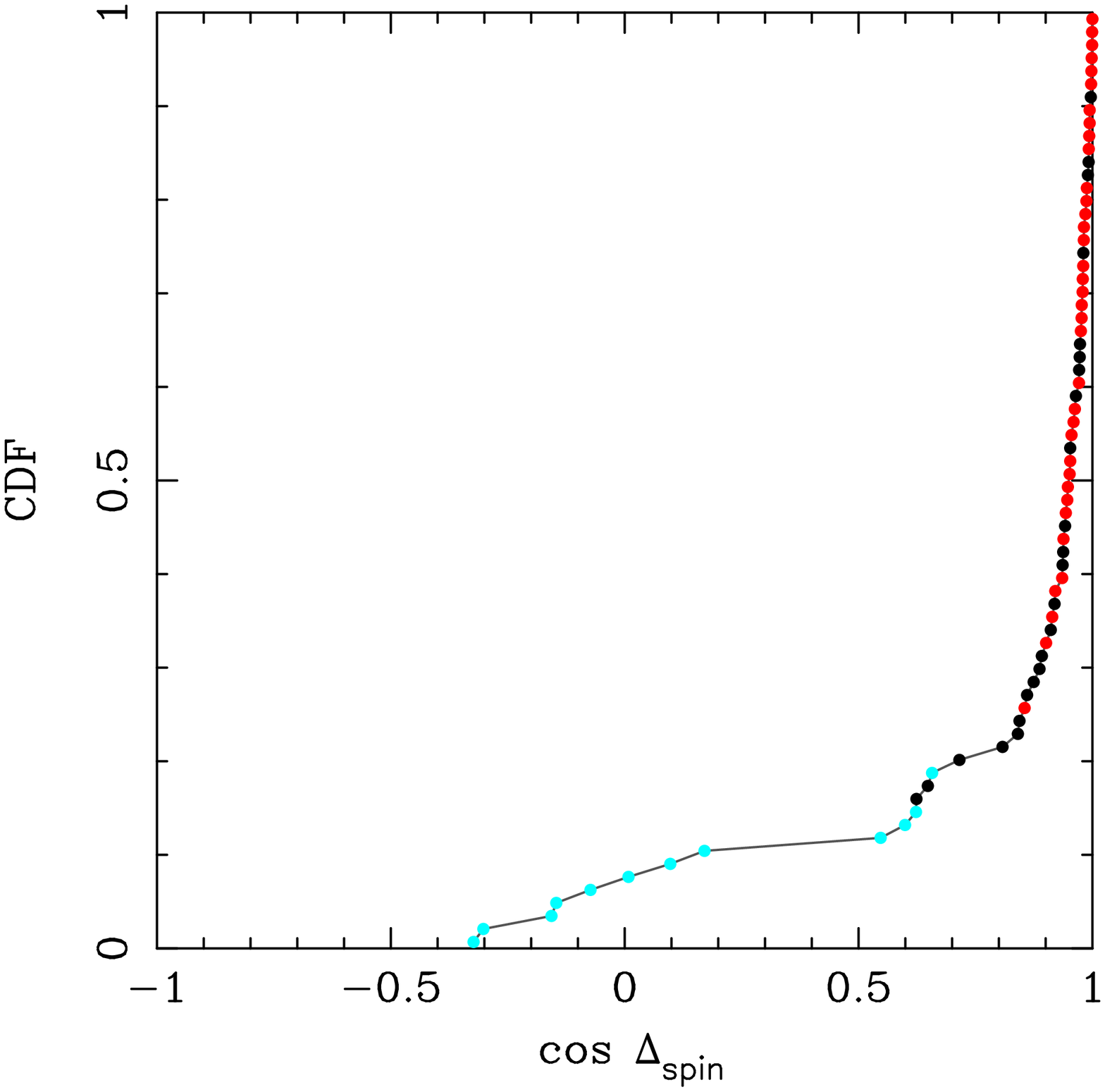}
    \caption{Cumulative distribution function for $\cos(\Delta_{\rm
      spin})$.  Color shows fit quality.
      \label{cdf_spin}}
  \end{minipage}%
  \hbox to 0.042\textwidth{}%
  \begin{minipage}[t]{\columnwidth}
    \includegraphics[clip=true,width=0.94\textwidth]{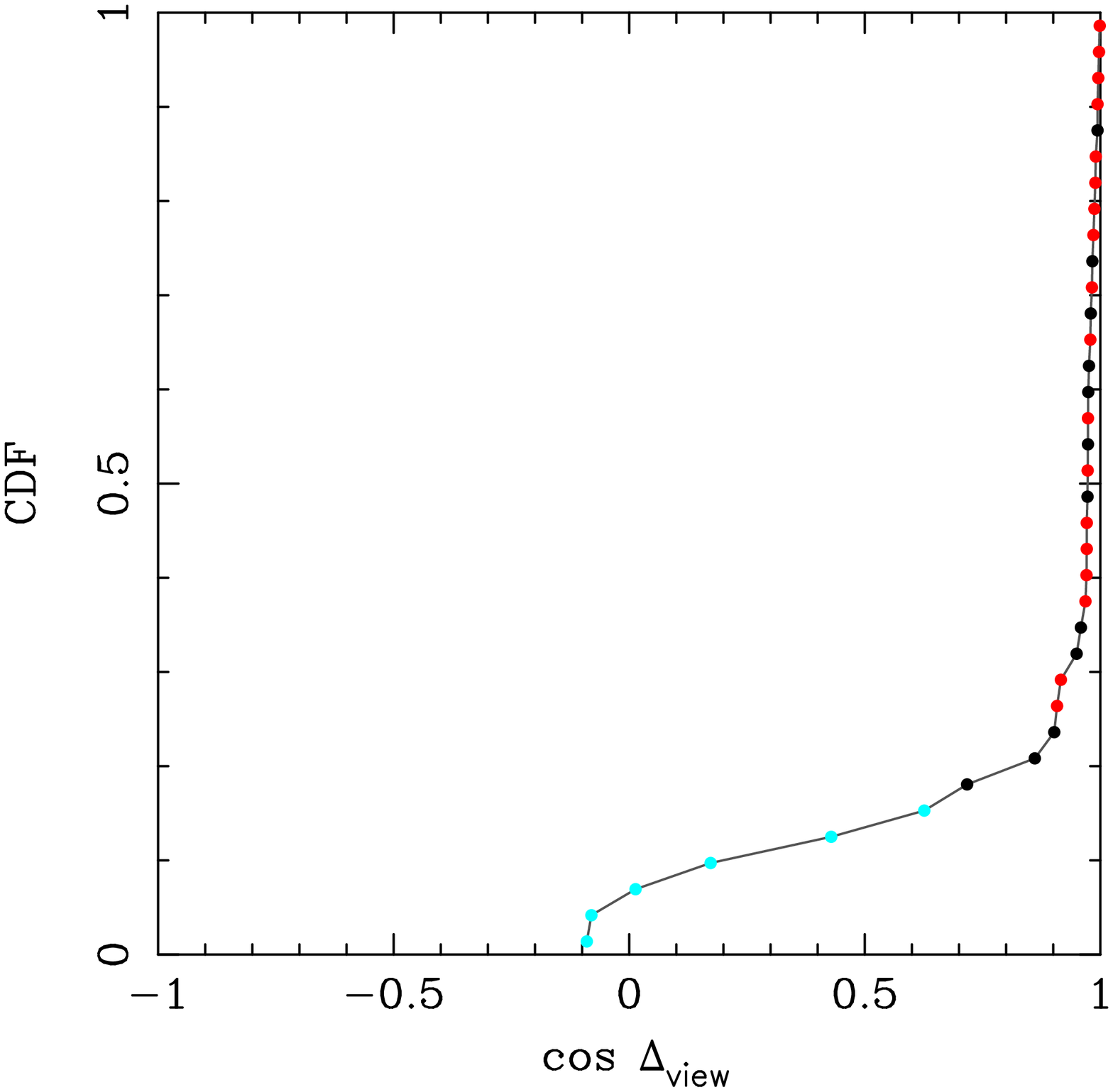}
    \caption{Cumulative distribution function for $\cos(\Delta_{\rm
      view})$.  Color shows fit quality.
      \label{cdf_view}}
  \end{minipage}%
\end{figure*}

To examine our overall accuracy in determining initial disk
orientations, we computed misalignments $\Delta_{\rm spin}$ between
true and fitted spin vectors for all $72$ disks.  Fig.~\ref{cdf_spin}
shows the cumulative distribution function of $\cos(\Delta_{\rm spin})
= \hat{\Vec{s}}_{\rm fit} \cdot \hat{\Vec{s}}_{\rm true}$; as in
previous plots, color indicates the subjective grade of each fit.  A
perfect match yields $\cos(\Delta_{\rm spin}) = 1$, while if
$\hat{\Vec{s}}_{\rm fit}$ was uncorrelated with $\hat{\Vec{s}}_{\rm
true}$ then $\cos(\Delta_{\rm spin})$ would be uniformly distributed
in the range $[-1,1]$ and the plotted points would fall along a
diagonal from lower left to upper right.  The actual distribution is
strongly peaked near $\cos(\Delta_{\rm spin}) = 1$, with the good fits
showing the smallest misalignments, the poor fits showing the largest
misalignments, and the fair fits falling in between.  This is
consistent with the previous figures, since $i$ and $\omega$ are just
angular coordinates for $\hat{\Vec{s}}$.  For the entire sample, the
median value is $\Delta_{\rm spin} = 18^\circ$, while for the $36$
disks in good fits the median is $\Delta_{\rm spin} = 12^\circ$.

In a similar fashion, Fig.~\ref{cdf_view} shows the distribution of
the misalignment in viewing direction, $\Delta_{\rm view}$, for all
$36$ fits.  Here $\cos(\Delta_{\rm view}) = \widehat{\Vec{Z}}_{\rm
fit} \cdot \widehat{\Vec{Z}}_{\rm true}$, perfect agreement again
yields $\cos(\Delta_{\rm view}) = 1$, and perfect ignorance would
distribute points along a diagonal from lower left to upper right.
Viewing direction is quite well determined; for the entire sample the
median $\Delta_{\rm view} = 13^\circ$, while for the $18$ good fits
alone the median is only slightly smaller, $\Delta_{\rm view} =
12^\circ$.  Note that an error circle with a radius of $12^\circ$
covers roughly $1$\% of the solid angle of a sphere; these fits are
{\it much\/} better than educated guesses!

\begin{figure*}[t!]
\begin{center}
\includegraphics[clip=true,width=0.45\textwidth]{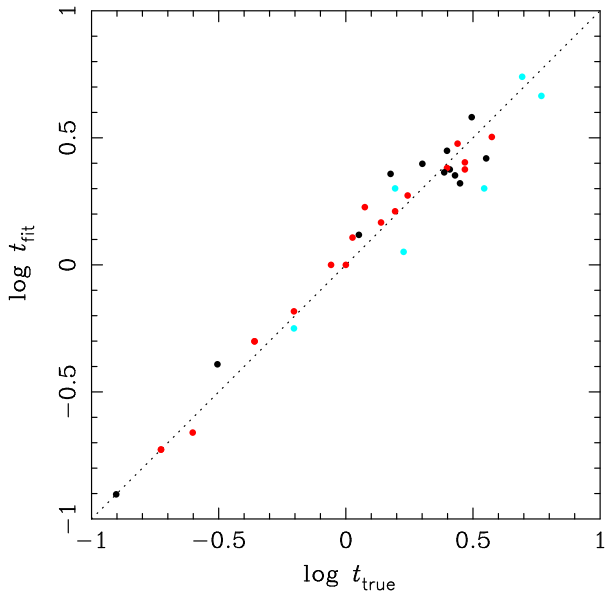}%
\hbox to 0.1\textwidth{}%
\includegraphics[clip=true,width=0.45\textwidth]{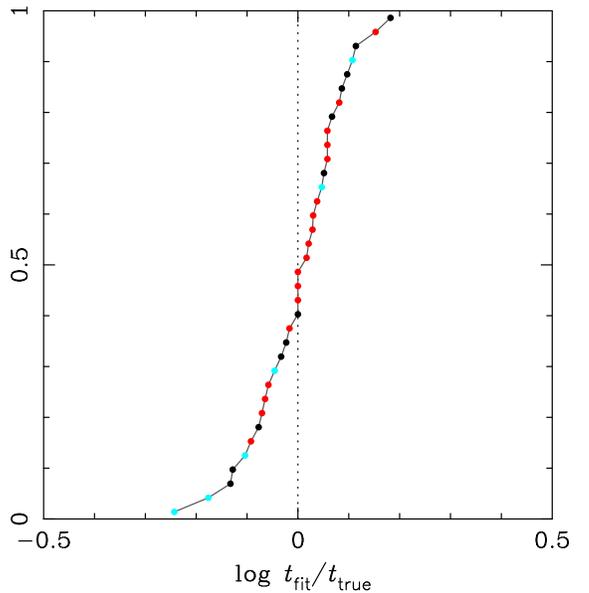}
\caption{Estimates of the dimensionless time since
  pericenter, $t$.  Left: scatter plot of $\log t_{\rm fit}$ against
  $\log t_{\rm true}$.  Right: cumulative distribution function of
  $\log t_{\rm fit}/t_{\rm true}$.  In both plots, color shows fit
  quality, and the dotted line represents perfect agreement ($t_{\rm
  fit} = t_{\rm true}$).
  \label{cmp_time}}
\end{center}
\end{figure*}

Fig.~\ref{cmp_time} compares the actual time $t_{\rm true}$ since
pericenter against the time $t_{\rm fit}$ obtained from the Identikit
fit.  The scatter plot on the left shows that the fitted and true
values are in good agreement, closely tracking each other throughout
the entire range of times; there is no evidence of bias or systematic
error, and the residuals appear to be random.  The plot on the right
shows the cumulative distribution of the fit/true ratio, $t_{\rm
fit}/t_{\rm true}$.  The symmetric appearance of this curve provides
further evidence that $t$ is accurately estimated by the Identikit
models.  Subjective fit quality appears to correlate with $t_{\rm
fit}/t_{\rm true}$; of the ten points at the two extremes of the
distribution, only one comes from a good fit.

\begin{figure*}[t!]
\begin{center}
\includegraphics[clip=true,width=0.45\textwidth]{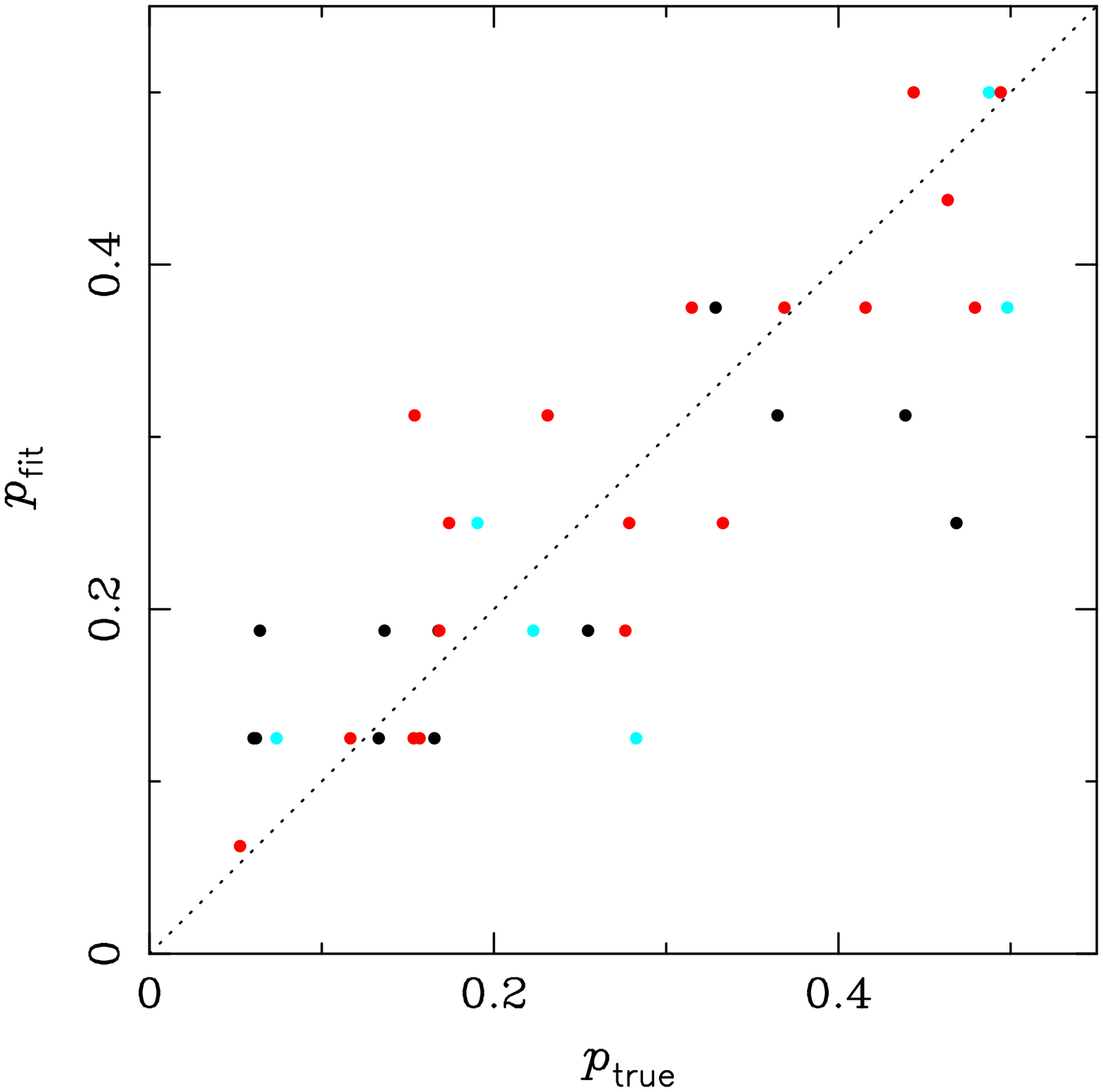}%
\hbox to 0.1\textwidth{}%
\includegraphics[clip=true,width=0.45\textwidth]{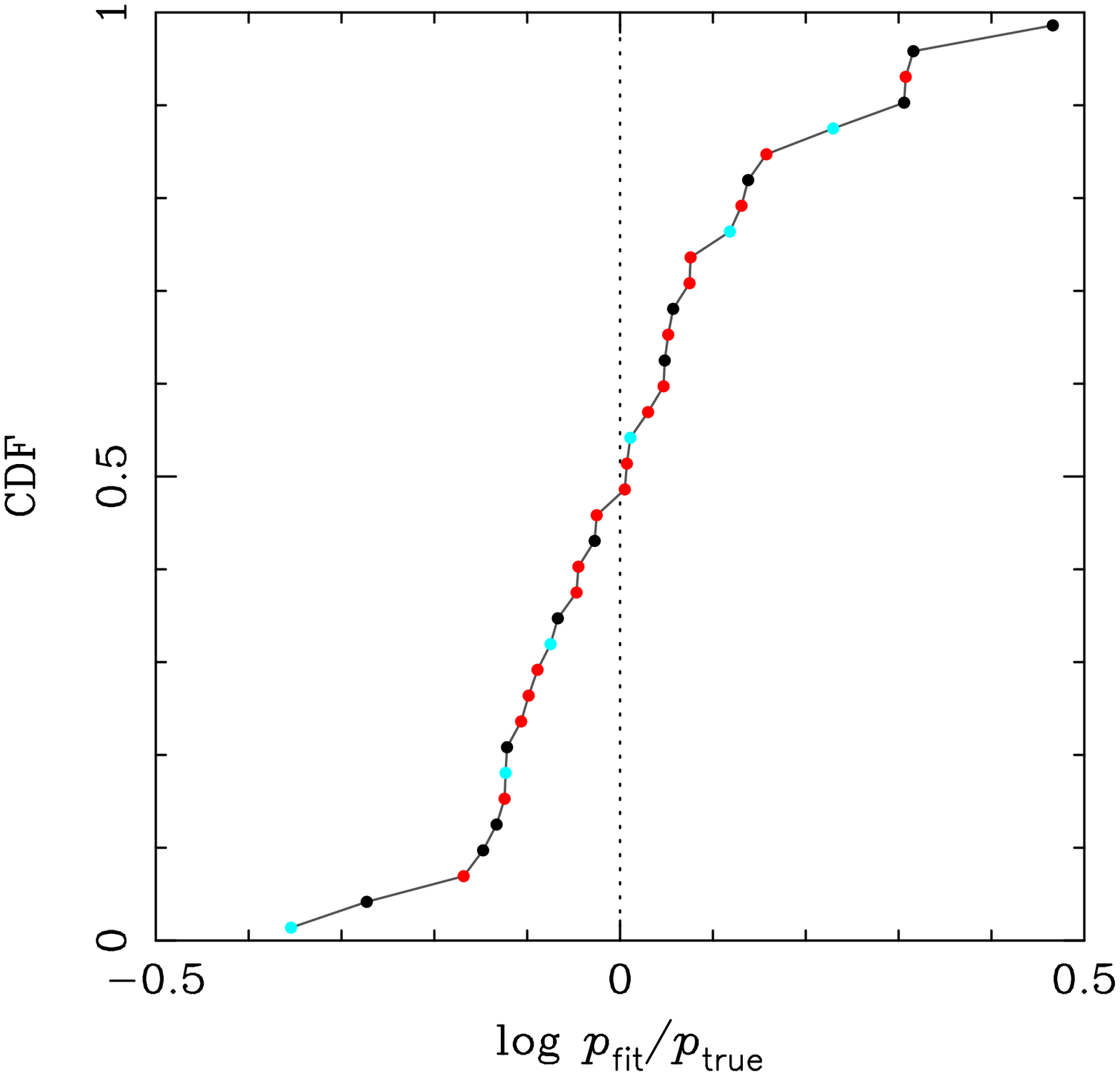}
\caption{Estimates of the dimensionless pericentric
  separation, $p$.  Left: scatter plot of $p_{\rm fit}$ against
  $p_{\rm true}$.  Right: cumulative distribution function of
  $p_{\rm fit}/p_{\rm true}$.  Colors and dotted lines as in
  Fig.~\ref{cmp_time}.
  \label{cmp_rperi}}
\end{center}
\end{figure*}

Fig.~\ref{cmp_rperi} compares fitted and true values of the
pericentric separation.  Here the range of $p$ values is rather small,
and the fact that $p_{\rm fit}$ can take on only eight discrete values
is evident.  There is a fair correlation between $p_{\rm fit}$ and
$p_{\rm true}$, although the points show considerable scatter.  The
cumulative distribution of $p_{\rm fit}/p_{\rm true}$ plotted on the
right is nonetheless fairly symmetric, and the median value of $p_{\rm
fit}/p_{\rm true}$ is very close to unity.  Curiously, there's not
much sign that the grades assigned the models correlate with $p_{\rm
fit}/p_{\rm true}$; fair and poor fits appear interspersed with good
ones throughout most of the distribution.

\begin{figure*}[t!]
\begin{center}
\includegraphics[clip=true,width=0.45\textwidth]{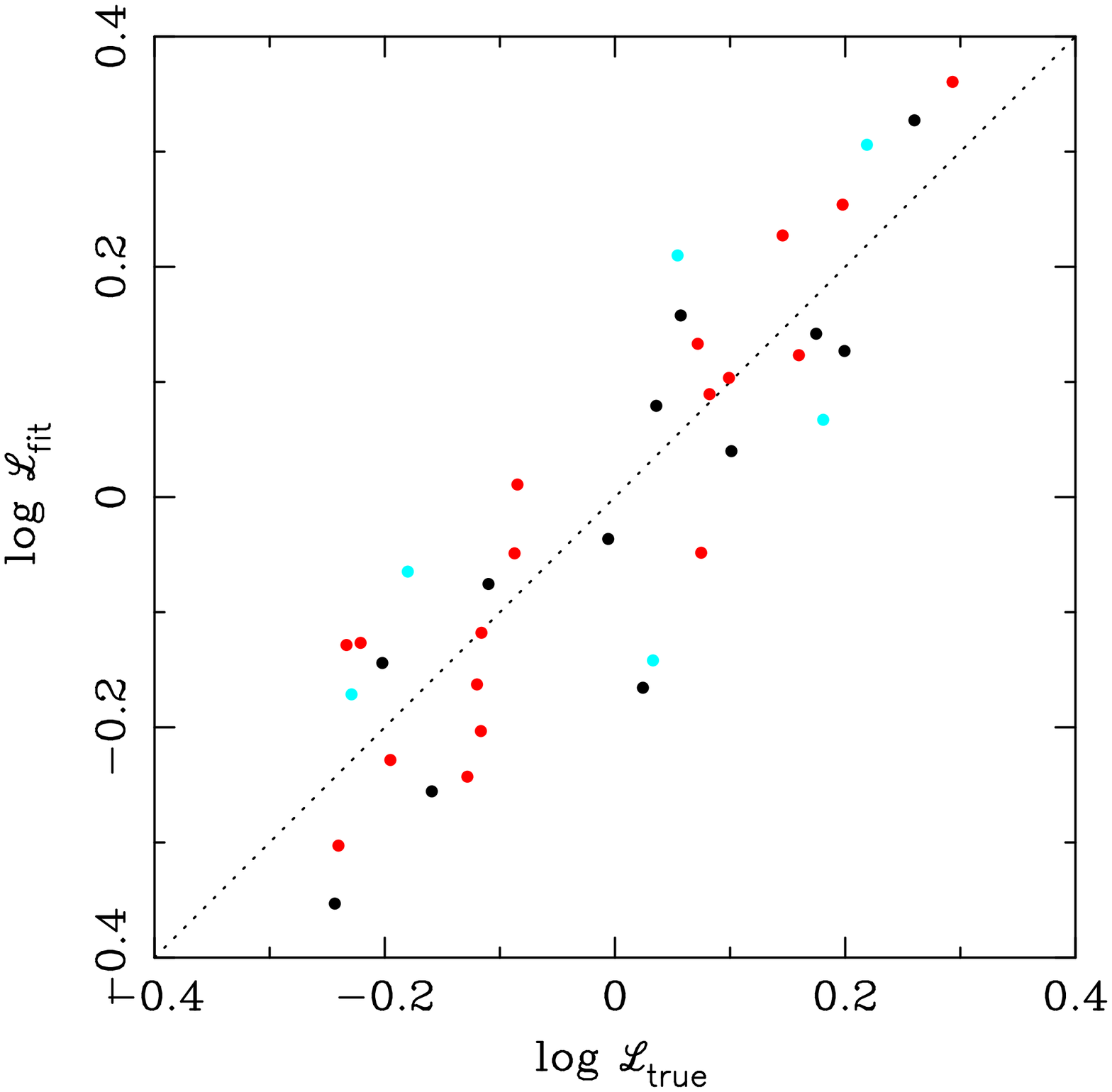}%
\hbox to 0.1\textwidth{}%
\includegraphics[clip=true,width=0.45\textwidth]{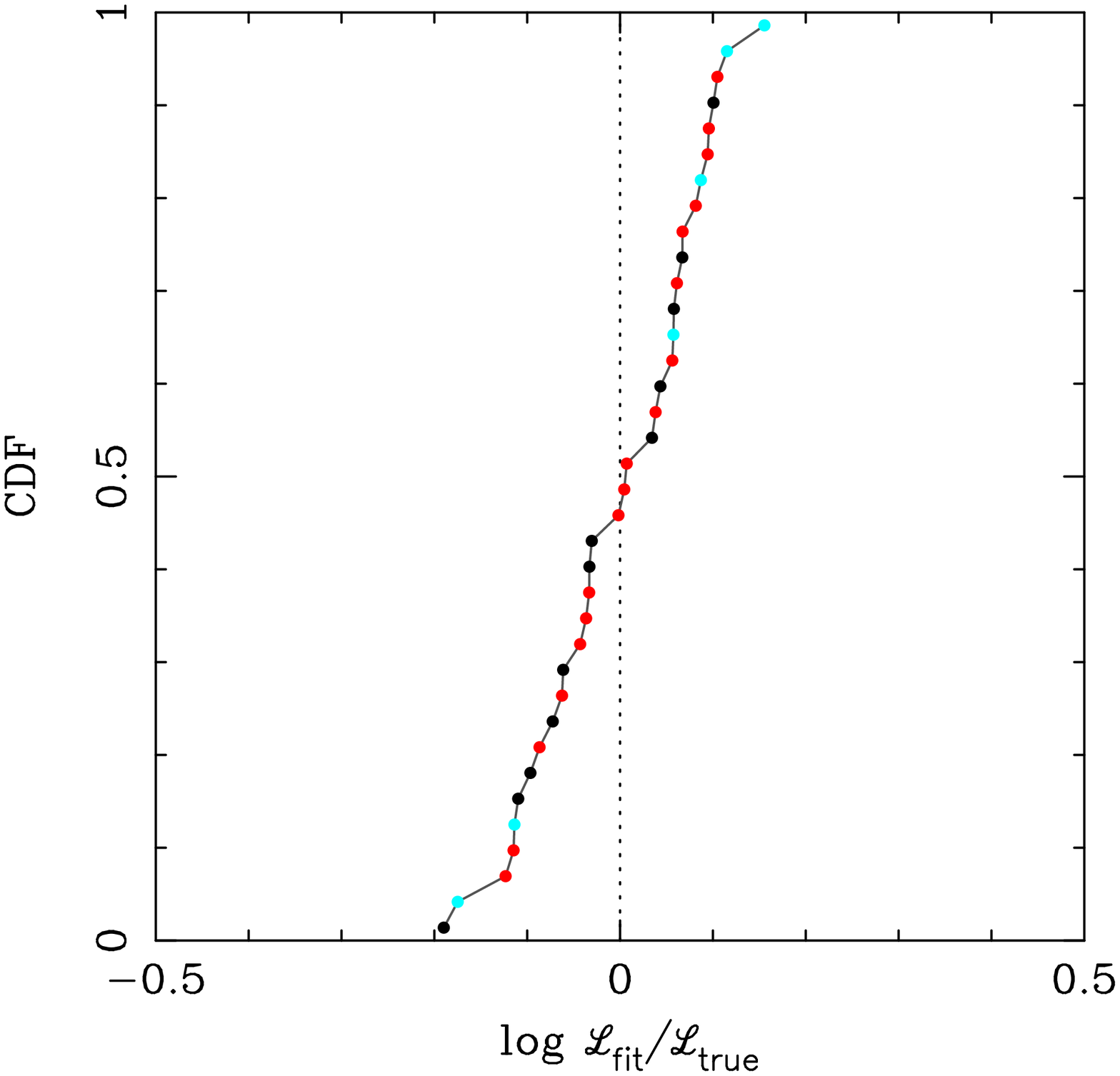}
\caption{Estimates of the length scale factor,
  $\mathscr{L}$.  Left: scatter plot of $\log \mathscr{L}_{\rm fit}$
  against $\log \mathscr{L}_{\rm true}$.  Right: cumulative
  distribution function of $\mathscr{L}_{\rm fit}/\mathscr{L}_{\rm
  true}$.  Colors and dotted lines as in Fig.~\ref{cmp_time}.
  \label{cmp_lscale}}
\end{center}
\end{figure*}

\begin{figure*}[t!]
\begin{center}
\includegraphics[clip=true,width=0.45\textwidth]{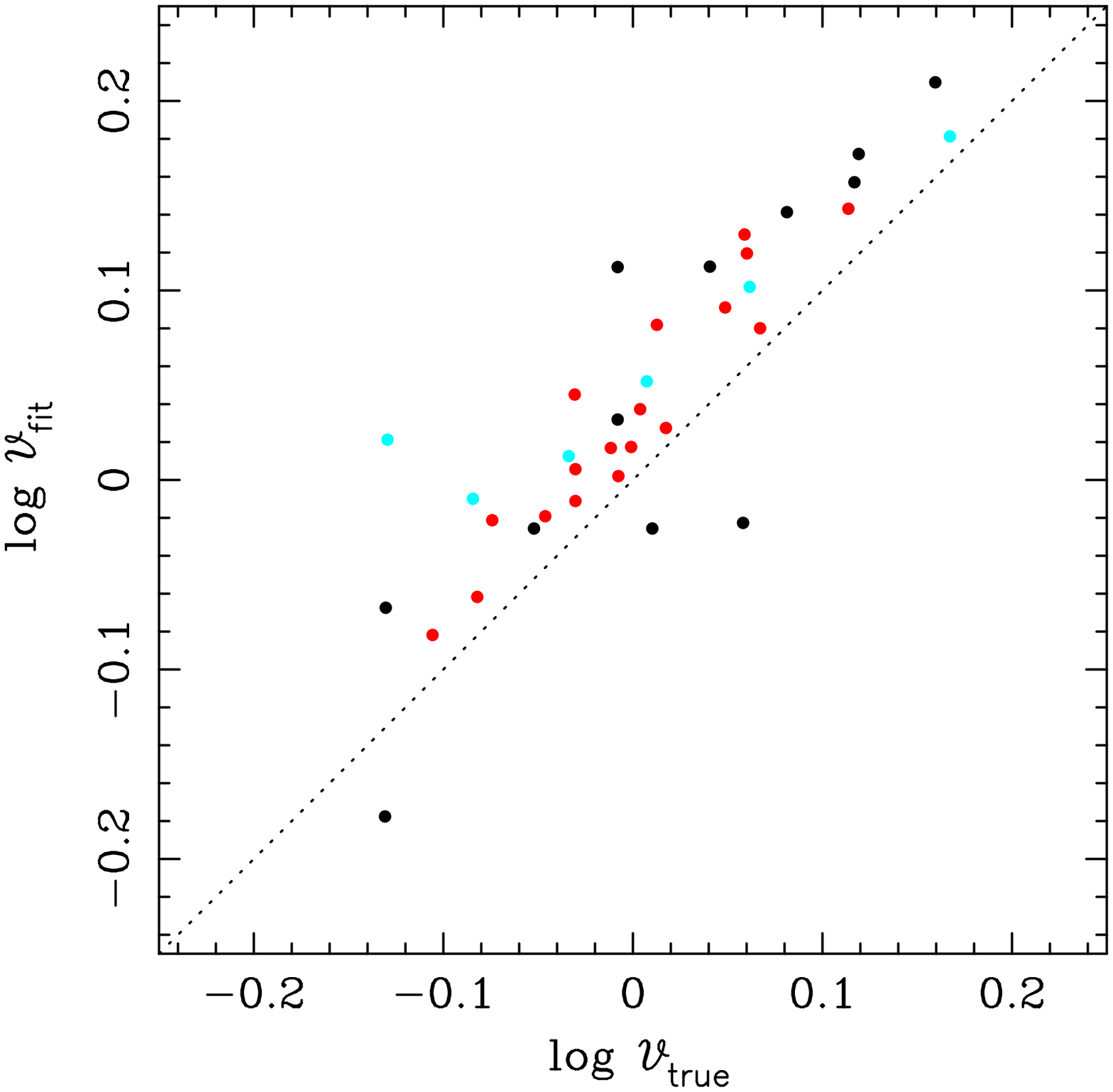}%
\hbox to 0.1\textwidth{}%
\includegraphics[clip=true,width=0.45\textwidth]{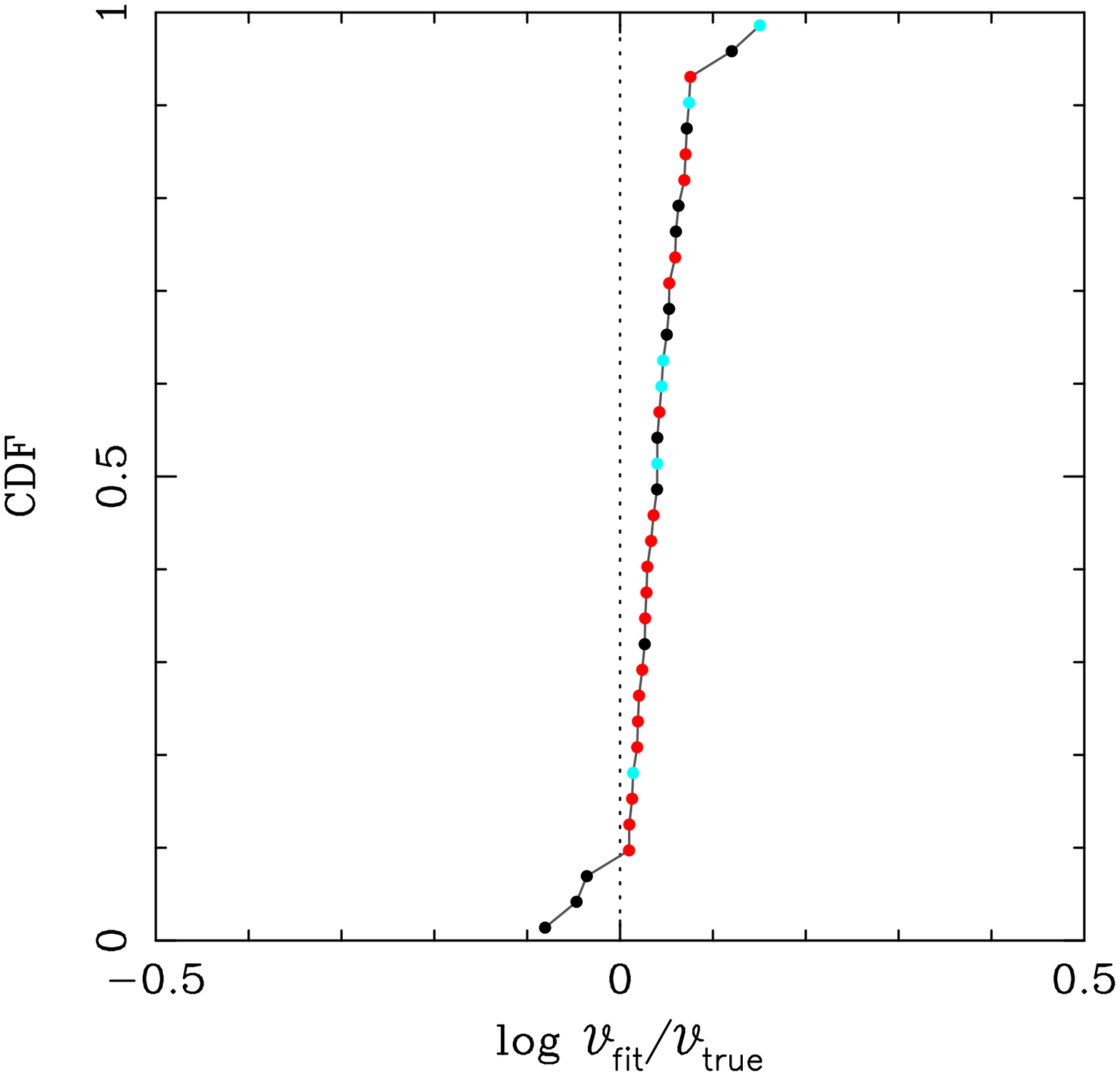}
\caption{Estimates of the velocity scale factor,
  $\mathcal{V}$.  Left: scatter plot of $\log \mathcal{V}_{\rm fit}$
  against $\log \mathcal{V}_{\rm true}$.  Right: cumulative
  distribution function of $\mathcal{V}_{\rm fit}/\mathcal{V}_{\rm
  true}$.  Colors and dotted lines as in Fig.~\ref{cmp_time}.
  \label{cmp_vscale}}
\end{center}
\end{figure*}

Identikit estimates of the length and velocity scale factors,
$\mathscr{L}$ and $\mathcal{V}$, are compared with their true values
in Figs.~\ref{cmp_lscale} and~\ref{cmp_vscale}, respectively.  The
length scale $\mathscr{L}$ is well-determined; the plot on the left
shows points scattered about the diagonal line, while the cumulative
distribution on the right shows a narrow spread with a median value of
$\mathscr{L}_{\rm fit}/\mathscr{L}_{\rm true}$ close to unity.  In
contrast, Identikit estimates of the velocity scale $\mathcal{V}$ show
a small but very definite bias; the distribution of $\mathcal{V}_{\rm
fit}/\mathcal{V}_{\rm true}$ is narrow, but clearly offset from unity.
Tests with isolated disks suggest that the absence of random motion in
the Identikit test-particle disks is responsible; see Appendix~C.

Table~\ref{fit_ratios} lists statistics for the ratios of fitted to
true values for $t$, $p$, $\mathscr{L}$, and $\mathcal{V}$; the
physical parameters $T$ and $P$ will be discussed in the next section.
Fit/true ratios for $t$ and $\mathscr{L}$ have fairly narrow
distributions centered on unity, confirming that these parameters are
determined accurately and without bias.  The fit/true distribution for
$p$, while somewhat broader, is also centered on unity.  The fit/true
distribution for the velocity scale factor $\mathcal{V}$ is quite
narrow, but the median value is $\sim 10$\% too high, showing again
that a small bias is present in fitting $\mathcal{V}$.

\subsection{Results: Residuals}
\label{sec:residuals}

\begin{figure}[t!]
\begin{center}
\includegraphics[clip=true,width=\columnwidth]{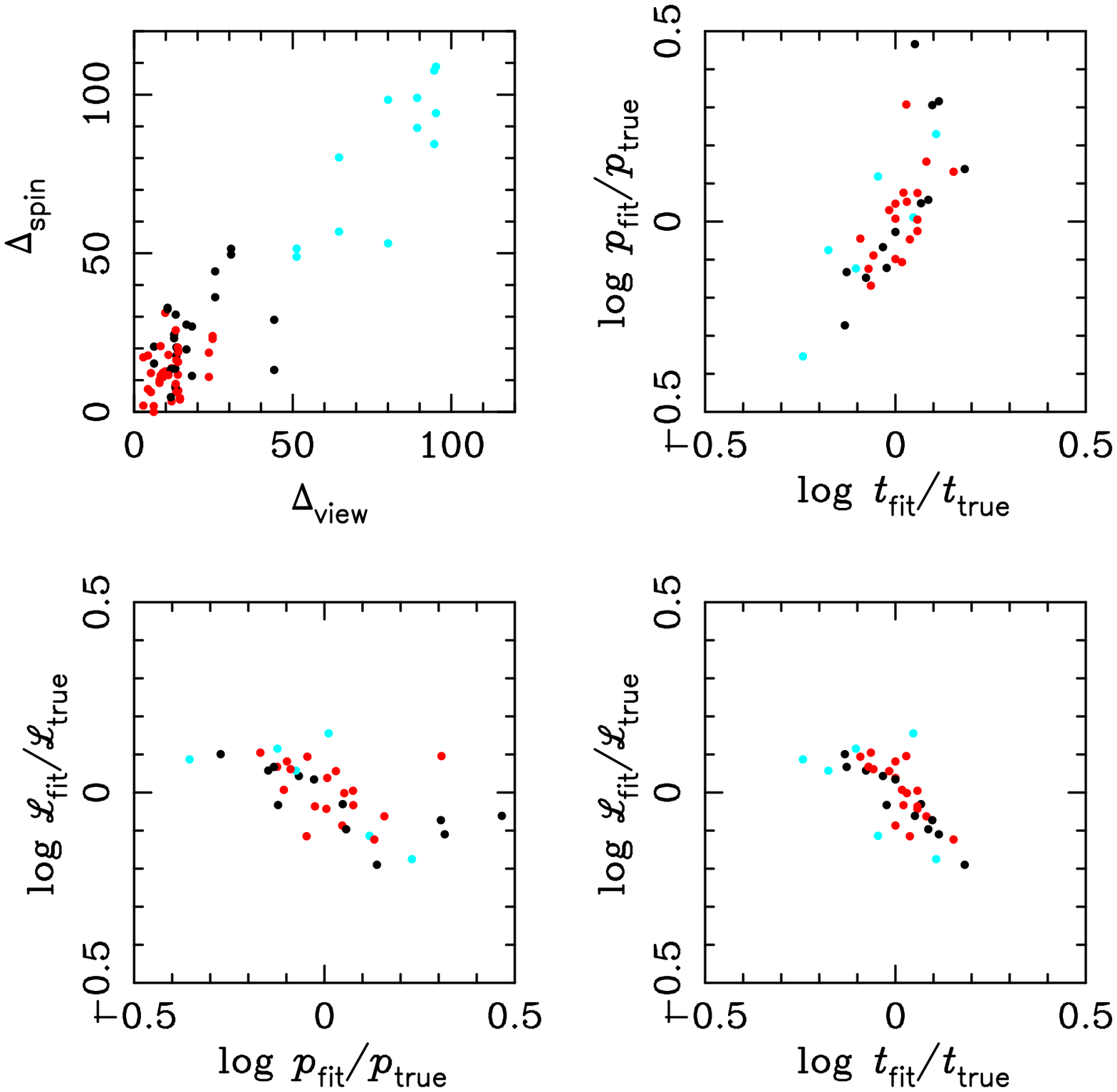}
\caption{Scatter-plots of statistically significant correlations
between residuals in Identikit models.  Color indicates quality of
fit.  The plot on the upper left contains twice as many points as the
others because $\Delta_{\rm spin}$ is plotted for each disk.
\label{residuals}}
\end{center}
\end{figure}

The errors determined by comparing the Identikit models with the
actual mergers define six independent residuals: viewing direction
($\Delta_{\rm view}$), spin direction ($\Delta_{\rm spin}$), time
since pericenter ($t_{\rm fit}/t_{\rm true}$), separation at
pericenter ($p_{\rm fit}/p_{\rm true}$), length scale
($\mathscr{L}_{\rm fit}/\mathscr{L}_{\rm true}$), and velocity scale
($\mathcal{V}_{\rm fit}/\mathcal{V}_{\rm true}$).  If each of these
residuals is plotted against the others, a total of fifteen ($6 \times
5 / 2$) potential relationships can be examined.  Most of these plots
show no measurable correlation; only the four in Fig.~\ref{residuals}
are significant (correlation coefficient $> 0.5$).

As shown in the upper left panel of Fig.~\ref{residuals}, errors in
disk orientation $\Delta_{\rm spin}$ are correlated with errors in
viewing direction $\Delta_{\rm view}$.  This correlation, which is
largely driven by the poor fits, is not hard to explain.  Once a
viewing direction has been selected, the next step is usually to
adjust the disk orientations; if clear guidance from tidal features is
lacking, the best one can do is to match each disk's position angle
and apparent tilt with respect to the line of sight.  A poor choice
for the viewing direction will induce comparable errors in spin
direction, as seen here.

The plot in the upper right panel
shows that residuals in the time since pericenter $t$ correlate with
residuals in pericentric separation $p$.  This correlation also has a
simple explanation; wider passages evolve and merge more slowly, so if
$t_{\rm fit}$ is for some reason overestimated during the matching
process then a larger $p_{\rm fit}$ can partly compensate for this
error.  As an extreme example, suppose $t_{\rm fit}$ was set so high
that an encounter with the correct $p_{\rm fit}$ would already have
merged by this time; by selecting a larger value for $p_{\rm fit}$,
the merger can be delayed until some time after $t_{\rm fit}$,
improving the match to the morphology.

A clear correlation between the residuals in $t$ and $\mathscr{L}$ of
the form $\mathscr{L}_{\rm fit}/\mathscr{L}_{\rm true} \simeq (t_{\rm
fit}/t_{\rm true})^{-1}$ is shown in the lower right panel of
Fig.~\ref{residuals}.  Such a correlation arises naturally in cases
where fits are strongly constrained by well-developed tidal tails.  In
a proper tail, material at the tip is moving fast enough to escape; at
late times this material will have asymptotically constant velocity
\citep{TT72}, implying that a tail's length grows in proportion to its
age \citep{S77}.  If $t_{\rm fit}$ is overestimated, tidal tails will
be longer in direct proportion, and a smaller value of
$\mathscr{L}_{\rm fit}$ will be needed to rescale them to their
correct physical lengths.  It's likely that similar considerations
also apply when other kinds of tidal features provide the primary
constraints; this is consistent with the results presented here, which
show that the good and fair fits fall close to $\mathscr{L}_{\rm
fit}/\mathscr{L}_{\rm true} \simeq (t_{\rm fit}/t_{\rm true})^{-1}$,
while the poor ones are more scattered.

Finally, we conjecture that the correlation between residuals of $p$
and $\mathscr{L}$ shown in the lower left panel is induced by the
$t_{\rm fit}/t_{\rm true}$--$\,p_{\rm fit}/p_{\rm true}$ and $t_{\rm
fit}/t_{\rm true}$--$\mathscr{L}_{\rm fit}/\mathscr{L}_{\rm true}$
correlations already described.  Of the three, the $p_{\rm fit}/p_{\rm
true}$--$\mathscr{L}_{\rm fit}/\mathscr{L}_{\rm true}$ correlation
shows the most scatter, and its outliers tend to match those in the
$t_{\rm fit}/t_{\rm true}$--$\,p_{\rm fit}/p_{\rm true}$ correlation.

An interesting consequence of these correlations emerges when the
dimensionless model parameters $t$ and $p$ are combined with the
length and velocity scale factors $\mathscr{L}$ and $\mathcal{V}$.  In
matching an Identikit model to observational data, $X$ and $Y$ might
be given in units of ${\rm kpc}$, and $V$ might be given in units of
${\rm km\,s^{-1}}$; the scale factors $\mathscr{L}$ and $\mathcal{V}$ would
then have units of ${\rm kpc}$ and ${\rm km\,s^{-1}}$, respectively, and
define a transformation from dimensionless model data to real physical
values.  The physical pericentric separation is $P = \mathscr{L} p$,
while the time since pericenter is $T = (\mathscr{L}/\mathcal{V}) t$.
A glance at the two lower plots in Fig.~\ref{residuals} suggests that
these correlations may actually {\it reduce\/} errors in estimates of
the physical parameters $T$ and $P$.  Table~\ref{fit_ratios} supports
this; the uncertainty in $P$, as indicated by the ratio of $3^{\rm
rd}$ to $1^{\rm st}$ quartile values, is about half the uncertainty in
$p$.  For $T$ the improvement is not as striking; the ratio of $3^{\rm
rd}$ to $1^{\rm st}$ quartiles is slightly smaller than the ratio for
$t$, but it appears that $\mathcal{V}$ introduces some additional
scatter and a small bias.

\subsection{Poor Matches}
\label{sec:poor_match}

For balance, it's worth taking a closer look at some of the less
successful matches.  We judged six out of $36$ of our solutions to be
poor fits to the ``observed'' data.  As already noted, two of the
systems with poor matches had very diffuse tidal features, while four
had fairly strong features.  Our difficulties in modeling the latter
are somewhat surprising; why, given the information which must be
present in strong tidal features, didn't the models turn out better?

\begin{figure*}[t!]
\begin{center}
\includegraphics[clip=true,width=0.30\textwidth]{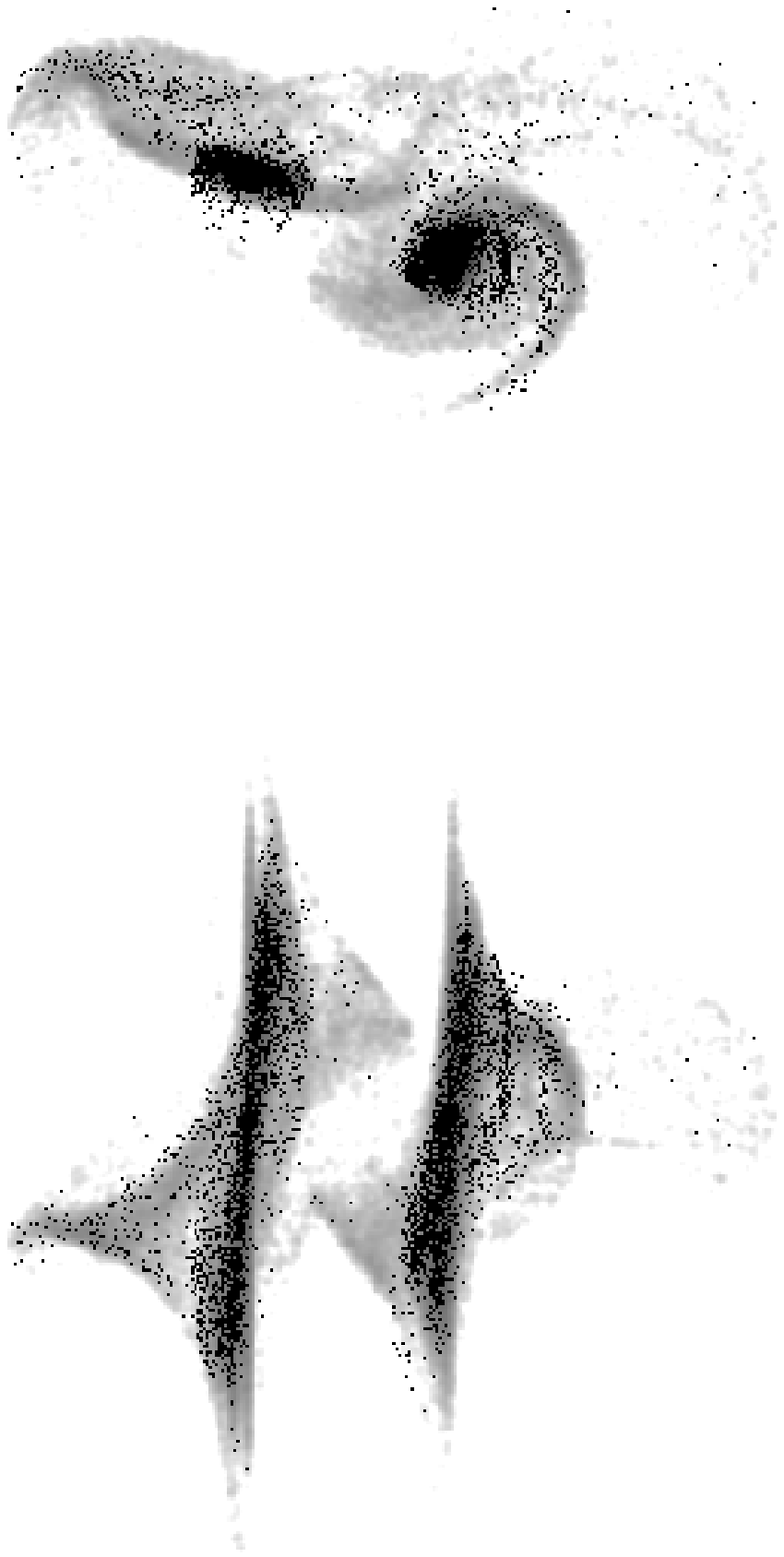}%
\hbox to 0.05\textwidth{}%
\includegraphics[clip=true,width=0.30\textwidth]{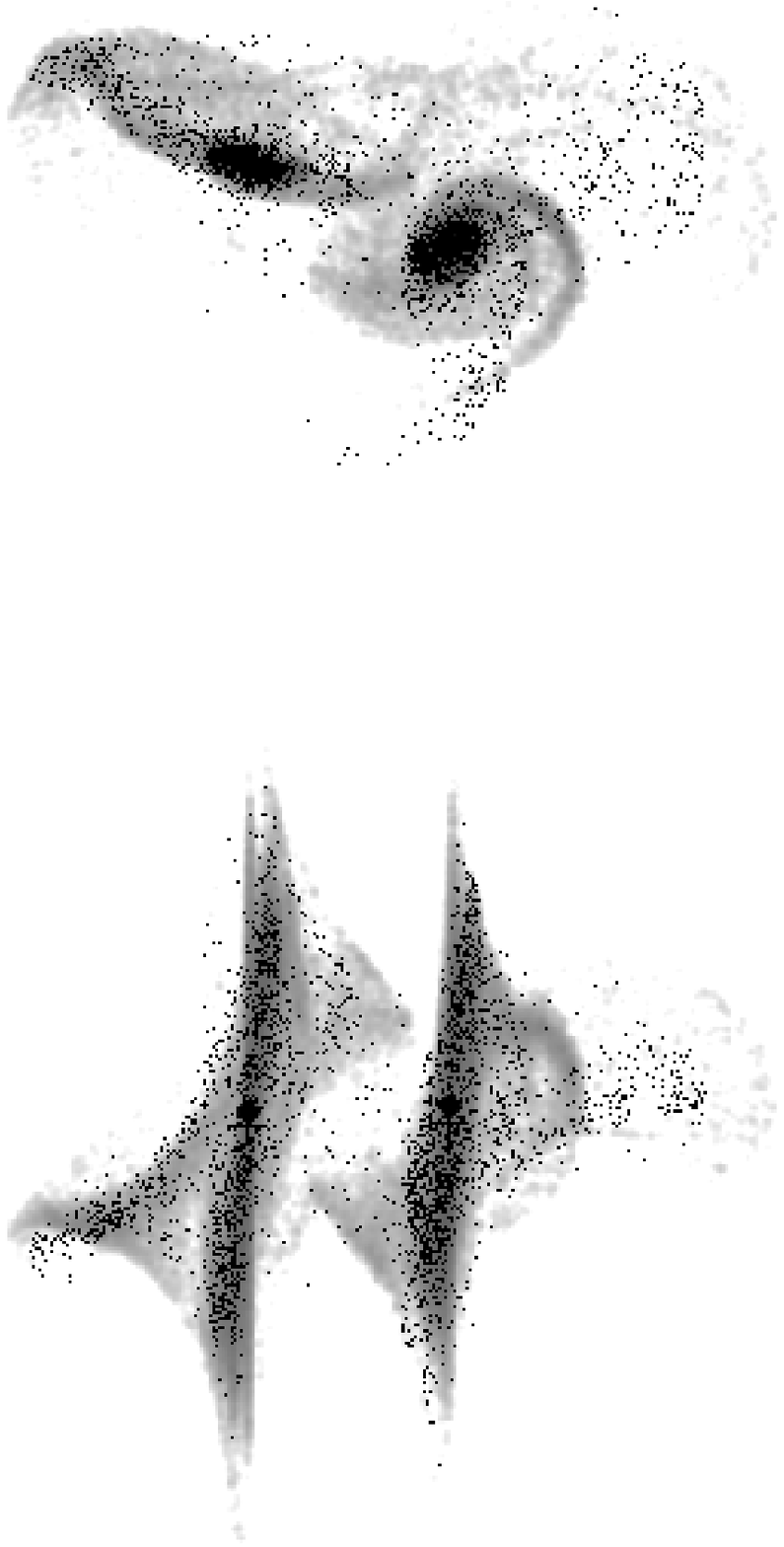}%
\hbox to 0.05\textwidth{}%
\includegraphics[clip=true,width=0.30\textwidth]{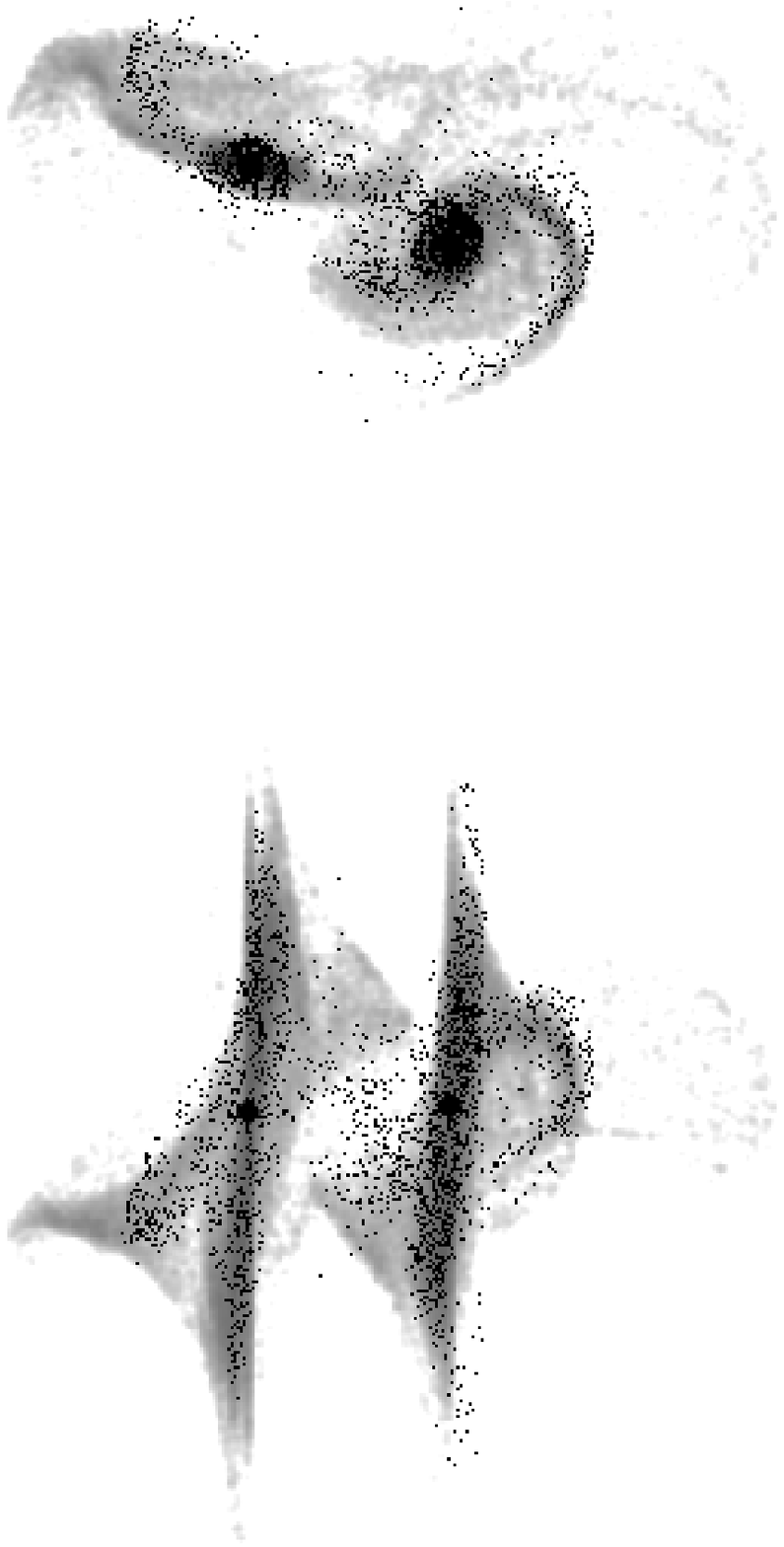}
\caption{Three models in search of a match.  Grey-scale images show
  the ``observed'' data (object~6 in Fig.~\ref{random_mergers}),
  while points represent models.  Top and bottom rows show projections
  on $(X,Y)$ and $(X,V)$ planes, respectively.  Left: the ``correct''
  Identikit solution.  Middle: a match to the shape and velocities of
  the more edge-on disk.  Right: a match to the spiral features of the
  more face-on disk.
  \label{poor_match}}
\end{center}
\end{figure*}

Fig.~\ref{poor_match} presents three versions of one of the more
difficult systems (object~6 in Fig.~\ref{random_mergers}).  For ease
of comparison, only the $(X,Y)$ and $(X,V)$ planes are shown; the
$(V,Y)$ views are harder to interpret since the galaxies partly
overlap.  The left-hand version shows an Identikit model with the
correct disk orientations, viewing direction, and velocity scale
``dialed in'' by hand.  Only a finite set of pericentric separations
are available in Identikit, so it was not possible to exactly match
the pericentric separation $p_{\rm true} = 0.223$; on the whole,
$p_{\rm fit} = 3/16$ gave a better match than $p_{\rm fit} = 1/4$
although the latter is numerically closer.  This closer approach
favored a somewhat different length scale ($\mathscr{L}_{\rm fit} =
0.644$ vs.\ $\mathscr{L}_{\rm true} = 0.590$) and a slightly earlier
time ($t_{\rm fit} = 1.62$ vs.\ $t_{\rm true} = 1.69$); both of these
adjustments are qualitatively consistent with the residual
correlations discussed in the last section.

The ``correct'' solution on the left in Fig.~\ref{poor_match} does a
poor job of matching the morphology (top).  It does better at matching
the kinematics (bottom), although some fairly large regions of phase
space are quite under-populated, and one of the spiral arms in the
more face-on disk produces a badly-placed streak of particles across
the $(X,V)$ projection.  By our subjective criteria, this solution is
a poor fit.  Since the disk orientations and viewing direction {\it
exactly\/} match those used to generate the grey-scale images, this
mismatch must be largely due to the lack of disk self-gravity in the
Identikit simulations.  In particular, the spiral morphology of the
more face-on disk is very poorly reproduced; with hindsight, we
recognize that this spiral probably owes its form to self-gravity.

The middle and right parts of Fig.~\ref{poor_match} present two of the
solutions we found while trying to match this system.  Essentially,
each matches one galaxy while failing to match the other.  The middle
solution does a plausible job of matching the morphology {\it and\/}
kinematics of the more edge-on disk, using a close passage ($p_{\rm
fit} = 1/16$) at an early time ($t_{\rm fit} = 1.12$).  Although it
does a very poor job with the more face-on disk, this solution is
actually not too bad -- the viewing angle is off by $47^\circ$, but
the disks are within $30^\circ$ of their correct orientations.  The
right-hand solution approximates the morphology and kinematics of the
more face-on disk, using a wider passage ($p_{\rm fit} = 3/16$) at the
same early time.  Ironically, although the decent-looking fit to the
face-on disk and somewhat plausible match to the edge-on disk led us
to adopt this solution as the best of a bad lot, it's actually a {\it
much\/} less accurate fit than the middle one -- the viewing angle is
off by $95^\circ$, while the face-on and edge-on disks are misaligned
by $109^\circ$ and $94^\circ$, respectively!  The splattering of
particles between the galaxies in the $(X,V)$ projection, populating a
region which should have been completely empty, was the critical flaw
which saved us from assigning a ``fair'' grade to this solution.

The other three systems which we failed to fit despite their strong
tidal features are a mixed lot.  In object~24, much as in the example
just discussed, the ``correct'' Identikit solution doesn't match the
morphology of either galaxy.  In object~11, the correct solution
matches one disk but fails to reproduce the morphology and kinematics
of its partner.  Finally, object~1 is reproduced quite accurately by
the correct Identikit model.  Our failure with object~1 was largely
due to insufficient patience, while with object~24 the mismatch was so
bad that the correct solution does not stand out when compared to
other possible fits.  Object~11 is intermediate; had we stumbled
across the correct match to one disk we might have recognized the
overall plausibility of this solution.  In sum, human limitations and
lack of self-gravity contribute about equally to the four poor matches
of systems with strong tidal features.

\section{Discussion}

Under somewhat idealized conditions, the Identikit methodology can
recover the key parameters of a galactic encounter from a single
data-cube.  If the method can also be applied to real data, it will be
a powerful tool for interpreting observations and reconstructing the
dynamical histories of interacting galaxies.  But before going
further, some limitations should be discussed.

\subsection{Limitations}

First, as noted above, Identikit simulations are not completely
self-consistent.  Halos and bulges are treated self-consistently, but
disks -- of necessity -- must be modeled with test particles.  Disk
structures requiring self-gravity, including bars and swing-amplified
spirals, will not be reproduced.  Consequently, these features can't
be used to match models to observations.  Moreover, orbital decay of
Identikit models is independent of disk orientation, whereas direct
passages are expected to decay faster than retrograde ones \citep{W79,
B92}.  For a pair of galaxies approaching their second passage, the
estimated time $t_{\rm fit}$ since first pericenter may be off by as
much as $\sim 20$\%; in applications requiring an accurate estimate of
$t_{\rm fit}$, Identikit models should be followed up with
self-consistent simulations.

Second, while the test particles we used to model disks are
collisionless, most kinematic tracers follow a specific phase of the
interstellar medium -- for example, H{\small I}, H$\alpha$, or CO.
Collisionless particles can approximate gas dynamics, but only if the
gas moves ballistically; streams of particles freely interpenetrate,
whereas gas will be deflected if it encounters shocks.  Examples
include mass transfer via genuine bridges formed in low-inclination
encounters \citep{TT72} and ``splash bridges'' due to hydrodynamic
forces in interpenetrating encounters \citep{S97}.  More subtle are
the modest star--gas offsets seen in low-inclination encounters with
extended gas disks \citep{M01}; it appears that the gas offset from
the stars does follow ballistic trajectories, while some gas initially
associated with the stars dissipates and falls back.  Additionally,
shocks may change the physical state of the gas; in particular, dense
molecular gas is often associated with material which has undergone
significant dissipation and hence {\it cannot\/} be modeled
collisionlessly.  But for most encounter geometries, gas in tidal
features should be well-approximated by collisionless particles;
exceptions can be recognized and allowed for in the fitting process.
Stars, especially populations pre-dating the onset of an encounter,
could be a useful complement to gas-phase tracers; absorption-line
spectroscopy would require prohibitively large amounts of telescope
time, but individual planetary nebulae are already providing kinematic
data for nearby systems \citep[e.g.,][]{HFFD95, DMFJC03}.

Third, two parameters were not included in the present experiment: the
orbital eccentricity $e$ and the mass ratio $\mu$.  In principle it's
straightforward to include these parameters in the fitting process,
although doing so increases the number of Identikit models needed.
Both parameters have a priori constraints -- orbits with $e \sim 1$
are favored theoretically (\citealt{TT72}; but see \citealt{KB06}),
while $\mu$ can sometimes be estimated photometrically assuming
constant $M/L$ ratios -- so it seemed reasonable to exclude them from
the initial experiments.  Putting these parameters in play would
probably increase the scatter in our fits; moreover, the three-way
correlation between residuals of $t$, $p$, and $\mathscr{L}$
(\S~\ref{sec:residuals}) might expand to involve residuals of $e$ and
$\mu$ as well.  Within the context of the models examined here,
there's scant reason to expect much cross-talk between these
parameters and others; in particular, disk orientations and viewing
angles are robustly constrained by tidal features and should be well
determined even if $e$ and $\mu$ are included in the fits.

Fourth, real galaxies have a range of rotation curve shapes,
reflecting a diversity of mass profiles \citep[e.g.,][]{CvG91,CGH06};
a single mass model is too limited.  The fact that we used the same
mass model for our artificial data {\it and\/} for the Identikit
simulations no doubt helped to reduce the uncertainties in our
solutions.  There's no reason why Identikit simulations cannot include
a variety of mass models, although this will increase the number of
choices to be made in modeling a galactic collision.  On the other
hand, if such simulations can discriminate between different mass
models then they would provide a way to analyze the structure of disk
galaxies.

\subsection{Previous Studies}

Models of interacting galaxies have a long history.  \citet{TT72}
presented test-particle models of four systems: Arp~295, M~51,
NGC~4676, and NGC~4038/9; the latter three have been revisited time
and again by other workers.  Kinematic information was initially
scarce and of uneven quality, so early modeling attempts focused on
reproducing the optical morphology.  Better velocity data has given
kinematics a more substantial role in more recent modeling efforts.
At the same time, faster computers and N-body algorithms have enabled
researchers to construct models incorporating self-gravity.

The methodology adopted by \citet{HM95} in their model of the
well-known merger remnant NGC~7252 includes several key practices
found in other successful models:

\begin{enumerate}

\item
Detailed velocity information, in the form of H{\small I} data
\citep{HGvGS94}, was available to constrain the model.

\item
Astrophysical arguments were used to estimate several critical
parameters -- specifically, the mass ratio $\mu$, the initial orbital
eccentricity $e$, and the pericentric separation $p$.

\item
The model focused on reproducing the large-scale and morphology and
kinematics of the tidal tails; these features evolved ballistically
since first passage and therefore carry a memory of the initial
encounter.

\item
Test-particle models, with rigid galaxy potentials constrained to
follow realistic merger trajectories, were used to narrow down the
range of parameter space.

\item
Fully self-consistent N-body models were used to refine the final
model.

\item
Simulation particles were plotted over orthogonal projections of the
data cube to show that the final model reproduced {\it both\/} the
morphology and the kinematics of NGC~7252.

\end{enumerate}

Table~\ref{dynamical_models} lists some interacting disk galaxies with
dynamical models incorporating significant kinematical constraints.
The progenitors of these systems span a range of mass ratios and
morphological types.  ``S+S'' systems involve two disk galaxies of
roughly comparable mass, both generally displaying significant tidal
features.  Most of these pairs are observed between first and second
passage; NGC~7252 is the only completed merger.
As \citet{SKBTEE05} note in modeling NGC~2207, earlier stages are
generally easier to fit.
In ``S+d'' encounters a disk galaxy is perturbed by a smaller
companion, while in ``E+S'' systems the disk is disturbed by an
elliptical of comparable mass.  Finally, in ``ring'' galaxies a
companion has plunged almost perpendicularly through a disk galaxy
\citep{ASM96}; these systems are relatively straightforward to model
since their geometry is fairly simple.

To varying degrees, the studies in Table~\ref{dynamical_models} all
followed the methodology used by \citet{HM95}.  Most had access to
detailed velocity information, usually obtained by H{\small I}
interferometry or H$\alpha$ Fabry-Perot imaging, although a few models
were based on long-slit spectroscopy.  Adopted orbital eccentricities
reflect a range of assumptions, not all equally plausible; for most
systems, orbits with $e \sim 1$ seem more likely since $e < 1$ begs
the question of what happened on the {\it previous\/} passage.  Both
test-particle and self-consistent techniques were used.  Not all
models were refined using fully self-consistent simulations; inasmuch
as orbital decay is critical for many of these systems, the use of
rigid potentials may be problematic in some cases.  

As the third column of Table~\ref{dynamical_models} shows, a wide
range of criteria were used to define an acceptable match to the
observations.  Many studies still seemed more focused on morphology
than kinematics, and less than half presented compelling quantitative
comparisons between models and data.  Matches labeled ``kin2'' used
2-D kinematical information, and presented the models and data in such
a way that direct and unambiguous comparisons could easily be made --
ideally, the model and data were overplotted, or at a minimum plotted
to the same scale and orientation.  Matches labeled ``kin1'' used 1-D
data (e.g., long-slit spectra or mean velocities plotted as functions
of a single coordinate), but again compared models and data directly.
A few studies, labeled ``gen'' (for genetic), evaluated matches
numerically; these will be discussed in \S~\ref{sec:genetic}.  Most of
the remaining studies, while drawing on spatially-resolved kinematic
information, presented essentially qualitative comparisons between
models and data; these matches are labeled ``qual'' in the table.
This designation is rather broad, ranging from studies which plotted
models and data on different scales to studies which matched general
kinematic trends or asserted, without providing quantitative evidence,
that the model matched the data.  Finally, a few studies which matched
morphology only are designated ``morph''; these are included when they
served as precursors to more comprehensive modeling efforts.

The present study, while restricted to artificial data, closely
parallels the approach of \citet{HM95}.  We depart from them in
treating the pericentric separation $p$ as a free parameter, and in
not using fully self-consistent simulations to finalize the models.
The latter, of course, is deliberate; one of our goals was to see if
an approach combining test-particle disks with self-consistent halos
can recover the encounter parameters of interacting disk galaxies.  In
practice, we envision using Identikit models to jump-start fully
self-consistent simulations.  We strongly concur that kinematic data
provide an acid test which any dynamical model must pass \citep{TT72,
BR91}, and that large-scale tidal features are the key to unlocking
the dynamical history of galactic encounters.  Finally, we emphasize
that direct and unambiguous comparison between the simulations and the
``observational'' data was a necessary ingredient of our approach.
Overplotting the particles on the data, as in Fig.~\ref{idkit_match},
is an effective way to present such comparisons.

\subsubsection{Genetic Algorithms}
\label{sec:genetic}

To date, most attempts to model interacting galaxies have relied on
expert judgement in selecting initial conditions and identifying good
matches between simulations and observations.  Recognizing the
considerable labour involved, several groups have tried to automate
the modeling process \citep{W98,TK01,GFP02}.  The proposed algorithms
have two essential components.  First, they must replace the
subjective comparison of the simulation particles ($\mathcal{P}$) and
the observed data ($\mathcal{D}$) with an objective criterion
$\mathcal{F}(\mathcal{P}, \mathcal{D})$ measuring goodness of fit.
Second, they must perform an efficient search of a very large
parameter space.  In view of the number of parameters involved, a
blind search is impractical; these groups have adopted strategies
mimicking biological evolution, generally known as {\it genetic
algorithms\/} \citep{H75}.

Genetic algorithms create a population of $N_{\rm pop}$ individuals,
each representing a possible solution to the problem at hand; in this
case, an individual defines a set of initial conditions and viewing
parameters.  The evolutionary fitness of individual $i$ is determined
by using its initial conditions and viewing parameters to produce a
particle distribution $\mathcal{P}_i$ which is evaluated using
$\mathcal{F}(\mathcal{P}_i, \mathcal{D})$.  Once all $N_{\rm pop}$
individuals have been evaluated, the fittest among them are bred
together to form a new generation, and the entire process is repeated.
After $N_{\rm gen}$ generations, the population converges toward a
nearly-optimal ensemble, with the fittest individual representing the
best approximation to the desired solution.

Genetic algorithms for modeling interacting galaxies have been tested
on artificial data \citep{W98,TK01,GFP02} and applied to real data for
NGC~4449 \citep{TK01} and NGC~5194/95 \citep{WD01,TS03}.  Typical
values of $N_{\rm pop} \simeq 10^2$ and $N_{\rm gen} \simeq 10^2$
imply that $\sim 10^4$ individuals must be evaluated to obtain a good
match; with test-particle methods, this can be done in a few hours of
CPU time.  The Identikit methodology (\S~\ref{sec:methodology}) could
be combined with a genetic algorithm, improving the treatment of
orbital decay and substantially reducing the CPU time required to find
a match.

However, the output of a genetic algorithm will be no better than the
evaluation function $\mathcal{F}(\mathcal{P}, \mathcal{D})$ used to
determine fitness.  The simplest approach is to coarsely grid
$\mathcal{P}$ on the $(X,Y)$ plane, and compare the result with an
equally coarse gridding of $\mathcal{D}$.  More recent implementations
incorporate velocity information as well, and there's no reason why
gridding can't be extended to 3-D $(X,Y,V)$ data.  But the first two
limitations of the Identikit method noted above also apply to existing
evaluation functions.  Features due to self-gravity in real systems
can be discounted by an expert when attempting to fit a test-particle
model, but may mislead an objective evaluation function, lowering the
fitness of accurate solutions.  And while H{\small I} is a good tracer
of kinematics, its distribution in tidal features is often quite
irregular; for example, a tail may appear as a series of clumps rather
than a connected structure.  An expert can recognize such tails as
connected structures, but an automatic procedure may reject solutions
which populate them with a smooth distribution of particles.  To
address these problems, recent genetic algorithm implementations
include routines for masking or weighting the observational data; it
remains to be seen if these techniques make genetic algorithms
competitive with human experts.

\subsection{Are Models Unique?}

Can a dynamical model reproducing the morphology and kinematics of an
interacting pair of galaxies be considered unique?  Claims to this
effect occasionally appear \citep[e.g.,][]{B88a,B88b,TK01}; skeptics,
paraphrasing John von Neumann\footnote{See \citet{D04} for one
version: ``with four parameters I can fit an elephant and with five I
can make him wiggle his trunk''.  John von Neumann could presumably
fit a whole herd of elephants with the parameters used to describe a
single galactic encounter!}, may be tempted to reply ``with that many
parameters I could fit an elephant''.  The $16$ parameters introduced
in \S~1 are all physically motivated and necessary to describe a
collision of two disk galaxies in 3-D; if a large number of parameters
per se was really a flaw, models of spectral line formation in stellar
atmospheres, requiring up to $\sim 90$ abundance parameters, would be
on shaky ground indeed!  Yet claims of uniqueness seem overconfident.
A particular match may be unique {\it within the universe of
possibilities defined by a given model\/}, implying that all the
parameters appearing in the model can be determined within reasonable
accuracy, and that no set of parameters outside this tolerance range
yield as good a match.  However, this is {\it not\/} the same thing as
uniquely determining the dynamical state of a pair of colliding
galaxies, which is specified by the distribution function
$f(\Vec{r},\Vec{v})$.

The problem of dynamically modeling isolated early-type galaxies,
which has generated an extensive literature, illustrates some of the
difficulties involved in determining $f(\Vec{r},\Vec{v})$.  Recent
studies \citep[e.g.,][]{vdVdZvdB08, vdBvdVVCdZ08} use Schwarzschild's
(\citeyear{S79}) method to fit models with equilibrium distribution
functions depending on three integrals of motion to stellar velocity
data obtained from integral field spectroscopy.  Such models are quite
successful at describing the orbital structure of galaxies and
diagnosing the presence of black holes and dark halos.  But not many
of these models are truly unique; the orientations and intrinsic
shapes of axisymmetric models appear uncertain \citep{KCEMdZ05,
vdBvdVVCdZ08}, while triaxial models pose additional difficulties
\citep{vdVdZvdB08}.  Determining $f(\Vec{r},\Vec{v})$ for one galaxy
is hard; doing so for a pair of galaxies seems harder still.

\begin{figure*}[t!]
\begin{center}
\includegraphics[clip=true,width=0.45\textwidth]{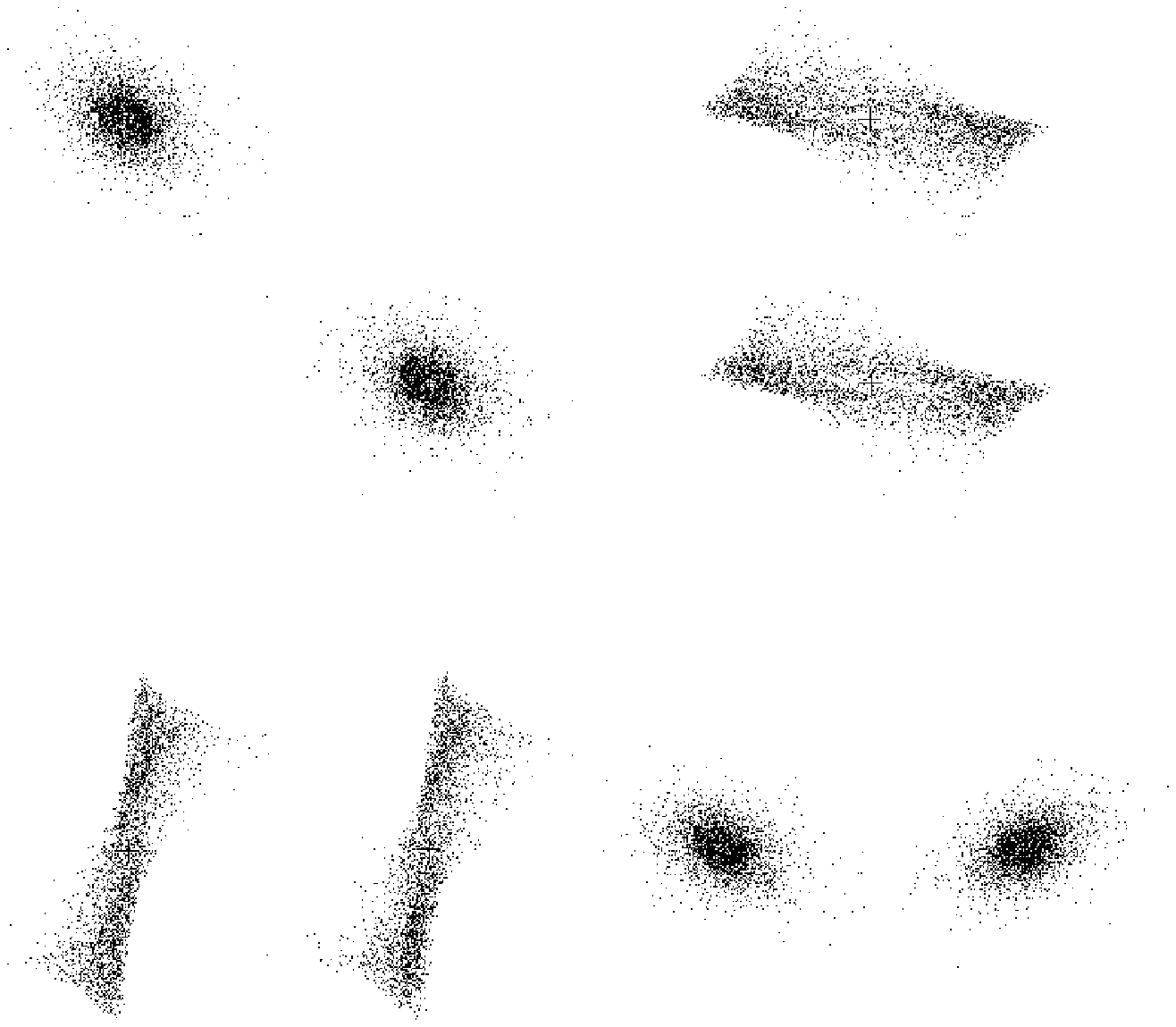}%
\hbox to 0.10\textwidth{}%
\includegraphics[clip=true,width=0.45\textwidth]{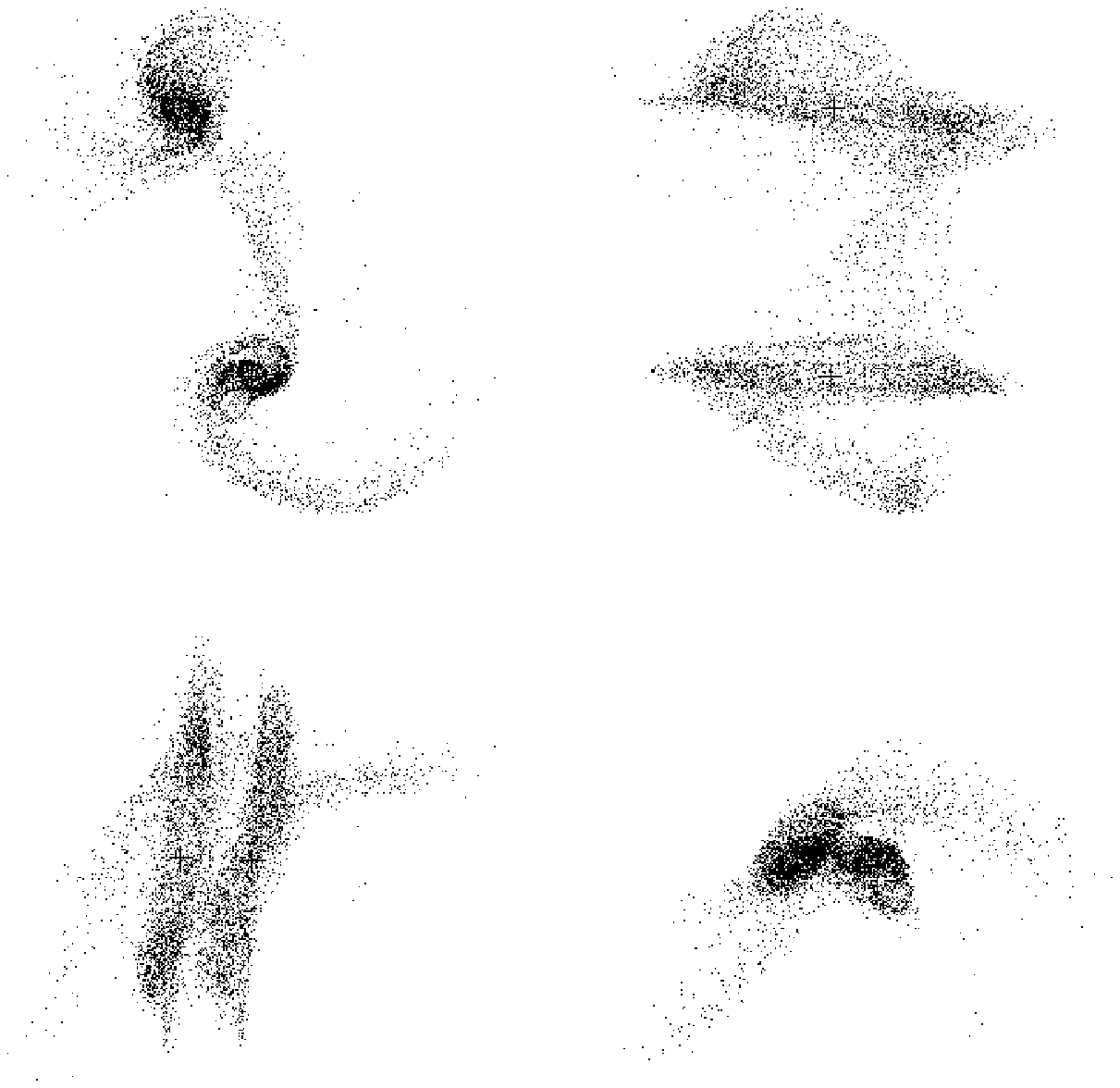}
\caption{Identikit simulation showing how a tidal encounter breaks
  degeneracy in disk orientations.  The system is viewed along the
  orbital axis.  The disks have inclinations $i_1 = 45^\circ$ and $i_2
  = 135^\circ$ and arguments $\omega_1 = \omega_2 = 30^\circ$.  Left:
  one time unit before pericenter; note that both disks present the
  same appearance on the $(X,Y)$, $(V,Y)$, and $(X,V)$ planes (top
  left, top right, and bottom left, respectively), although the
  $(X,Z)$ projection (bottom right) shows that they do not have the
  same orientation in space.  Right: one time unit after pericenter;
  the difference between the two disks is now evident in all four
  projections. 
  \label{degeneracy}}
\end{center}
\vspace{0.5cm}
\end{figure*}

In the context of the model described here, it may not even be obvious
that we can hope to constrain even the most basic parameters.  The
orientation of a single disk, unless it happens to be exactly edge-on,
cannot be determined from a data cube $F(X,Y,V)$; for example, a disk
tilted by $45^\circ$ to the line of sight produces {\it exactly\/} the
same data cube as one tilted by $135^\circ$.  (This ambiguity can be
resolved by using dust lanes to determine which side of the disk is
closer, although doing so goes beyond a strictly kinematic approach to
galaxy modeling; see, e.g., the discussion in \citealt{SKBTEE05}.)  By
extension, a pair of disks yield a four-fold degenerate solution,
since each disk has two possible orientations.  However, tidal
interactions between disks break this degeneracy; as
Fig.~\ref{degeneracy} shows, two disks which initially display
identical $F(X,Y,V)$ distributions can be differentiated {\it after\/}
an encounter.  In this example, as in most of our Identikit solutions,
the disk orientations $(i_1,\omega_1)$ and $(i_2,\omega_2)$ are
strongly constrained; a little ``wiggle room'' exists around the
actual values, but no radically different choice can reproduce the
``observed'' morphology and kinematics.

Taken together the results in \S~\ref{sec:parameters} show that almost
all of our $18$ ``good'' and $12$ ``fair'' fits accurately constrain
the disk orientations, viewing direction, time since pericenter,
pericentric separation, and scale factors.  We did not examine errors
in center-of-mass position and velocity, but the offsets in our fits
were always very small, so presumably these parameters are also
well-constrained.  While some solutions are better than others, all
$30$ of these fits appear to be unique; the estimated parameters are
always in the neighborhood of the actual values, and often very close
indeed.

The $6$ ``poor'' fits tell a different story.  Object~16 (see
Fig.~\ref{random_mergers}), a wide retrograde encounter ``observed''
long after pericenter, has such attenuated tidal features that fitting
this system is much like fitting a pair of disks {\it before\/} their
first encounter; our fit is clearly degenerate.  As
\S~\ref{sec:poor_match} describes, object~6 also produced non-unique
fits.  Other poor solutions may also be degenerate, in view of the
large misalignments in viewing direction and disk orientation they
display (see Figs.~\ref{cdf_spin} and \ref{cdf_view}).

The fact that we were able to produce well-constrained and reasonably
accurate solutions for $30$ out of $36$ systems depends critically on
the quality of the ``observational'' data we used.  Data with lower
resolution and added noise would yield a less favorable outcome; in
particular, incomplete velocity information can seriously compromise a
solution.  Models of NGC~7252 illustrate this point; early attempts
using very limited velocity data concluded that this object resulted
from a merger of two {\it retrograde\/} disks \citep{BR91}, but once
high-quality H{\small I} data was available a very convincing solution
involving a collision of two {\it direct} disks was soon found
\citep{HM95}.  Fortunately, instruments like the Expanded Very Large
Array now coming on-line should make it far easier to obtain the
detailed velocity information needed to accurately constrain dynamical
models of interacting galaxies.

However, even a surfeit of data will not guarantee an unambiguous
model in every case.  One example may be the ``Whirlpool'' galaxy,
NGC~5194/95; after three and a half decades of modeling based on
ever-better H{\small I} data there is still no consensus regarding the
number of passages required to account for NGC~5194's optical
morphology {\it and\/} extended H{\small I} tail \citep[e.g.][and
references therein]{DMFJC03}.  Models in which NGC~5195 makes only one
passage can account for many of the observations but have trouble
matching the tail velocities, while models in which this galaxy makes
{\it two\/} passages match the velocities better but yield double
tails \citep[see][Fig.~4]{SL00}.  One possible solution to this puzzle
is to assume that NGC~5194's H{\small I} disk was warped even {\it
before\/} its first and only encounter with NGC~5195; this could tilt
the tail with respect to the inner disk in such a way as to match the
observed velocities.  Of course, pre-existing warps vastly complicate
questions of uniqueness!  In the specific case of NGC~5194/95,
stronger evidence in favor of a single passage seems necessary to
justify introducing such a warp; better data on {\it stellar\/}
velocities may help \citep{DMFJC03}.

Our limited knowledge of dark matter is a more fundamental barrier to
the creation of truly unique models for interacting galaxies.  Models
of isolated galaxies in equilibrium can represent the dark matter as a
potential imposed by hand.  In contrast, models of interacting systems
should really treat dark matter as a full participant in the dynamical
equations\footnote{To be sure, not every study in
Table~\ref{dynamical_models} did this, but the speed of modern
computers leaves little excuse for not treating the dark matter
self-consistently in future work!}.  Given how little we know about
dark matter, a good deal of guess-work will be part of any such model,
and this is unlikely to change unless we can somehow measure the
detailed distribution {\it and\/} kinematics of dark matter in
individual galaxies.  Nonetheless, models of interacting galaxies have
already set limits on the radial structure of dark halos \citep{DMH96,
MDH98, B99, SW99}, and further studies may yield additional insights
into the nature, distribution, and dynamics of dark matter.

While it may never be possible to determine a unique distribution
function $f(\Vec{r},\Vec{v})$ for a specific pair of interacting
galaxies, there are still compelling reasons to construct detailed
models.  A good model of a galactic collision can serve as a unifying
hypothesis, providing a context to integrate and interpret a variety
of observations; for example, a time-line of past encounters may help
to understand interaction-induced star formation.  Conversely, a bad
model which contradicts other lines of evidence can be rejected.
Viewed as a means to an end, dynamical modeling stands to teach us a
good deal about galactic encounters and galaxies in general.

\subsection{Extensions}

While the current Identikit system is essentially an interactive
modeling tool, the basic approach offers other possibilities.  For
example, once the orbital parameters $p$, $e$, and $\mu$, time $t$,
viewing angles $(\theta_{\rm X},\theta_{\rm Y},\theta_{\rm Z})$, and
scale factors $\mathscr{L}$ and $\mathcal{V}$ have been fixed, it is
possible to invert the mapping from initial disk orientation
$(i,\omega)$ to the data cube $F(X,Y,Z)$, and ask which $(i,\omega)$
values allow a disk to populate a given point in the $(X,Y,V)$ space.
As a rule, each $(X,Y,V)$ point corresponds to some range of
$(i,\omega)$ values, but {\it different\/} points along one tidal
structure should all be populated by the {\it same\/} disk, so this
range can be constrained.  This approach could effectively automate
the process of finding disk orientations once other parameters have
been selected; it could particularly useful in establishing confidence
limits for key parameters once an initial match has been determined.

The Identikit technique has some interesting similarities to
Schwarzschild's (\citeyear{S79}) method for building triaxial
equilibrium systems.  In a nutshell, Schwarzschild started with a
stationary mass model, populated it with all possible orbits, and
figured out how to add up the time-averaged density distributions of
these orbits so as to recover the original mass model.  We start with
a time-dependent mass model generated by an encounter of two spherical
systems; each sphere is initially populated with all possible circular
orbits.  But instead of finding a weighted sum of these orbits which
reproduces the mass distribution, we try to select one co-planar
family of orbits from each sphere to match the observed kinematics and
morphology of interacting galaxies.  Perhaps some variant of the
algorithms used in Schwarzschild's method to determine orbital weights
could be applied to the problem of modeling interacting galaxies.

We hope to implement some of these extensions in Identikit~2.

\acknowledgements

J.E.B. thanks Fran\c{c}ois Schweizer, Breannd\'an Nuall\'ain, Shin
Mineshige, and Piet Hut for valuable conversations and comments, and
the Observatories of the Carnegie Institute of Washington, the
California Institute of Technology, Kyoto University, and the Japan
Society for the Promotion of Science for support and hospitality.  We
thank Curtis Struck for an open and constructive referee report.  The
National Radio Astronomy Observatory is a facility of the National
Science Foundation operated under cooperative agreement by Associated
Universities, Inc.

\appendix

\section{Identikit Model Construction}

To set up initial conditions for Identikit models we need to construct
spherical equilibrium N-body models with mass profiles $m(r)$ given by
(\ref{eq:total-mass-prof}).  We use Eddington's (\citeyear{E16})
formula \citep[e.g.,][p.~236]{BT87} to compute the distribution
function $f(E)$.  \citet{KMM04} used this approach to construct
equilibrium halos with density profiles described by a three-parameter
family of models \citep[e.g.,][]{Z96}.  We go one step further by
correcting for the finite resolution (i.e., ``softening'') of the
N-body force calculation.

We account for the effect of Plummer softening \citep{A63, HB90} in
the N-body simulations by introducing a quasi-empirical transformation
of the total mass profile (Barnes, in preparation):
\begin{equation}
  \overline{m}(r) =
    \left[
      1 + (2/3)^{(\kappa/\alpha)} (\tilde{\epsilon}/r)^{\kappa}
    \right]^{(\alpha/\kappa)} \, m(r) \, .
  \label{eq:smooth-mass-profile}
\end{equation}
where $\alpha$ is the logarithmic derivative of the density profile as
$r \to 0$, the parameter $\tilde{\epsilon}$ is comparable to the
softening length $\epsilon$, and the parameter $\kappa$ adjusts the
shape of the transition near $r \sim \tilde{\epsilon}$.  This smoothed
mass profile is then used to compute the potential:
\begin{equation}
  \frac{d\Phi}{dr} = G \frac{\overline{m}(r)}{r^2} \, ,
  \label{eq:potential-equation}
\end{equation}
where $\Phi \to 0$ as $r \to \infty$.  After expressing the original
density profile $\rho(r) = (4 \pi r^2)^{-1} d m / d r$ as a function
of this potential $\Phi$, we compute the distribution function:
\begin{equation}
  f(E) = \frac{1}{\sqrt{8} \pi^{2}} \, \frac{d}{dE}
    \int_{E}^{0} d\Phi \, (\Phi - E)^{-1/2} \, \frac{d\rho}{d\Phi}
  \label{eq:eddington-formula}
\end{equation}

Once $f(E)$ has been calculated, generating an N-body realization with
$N_{\rm sphr}$ massive particles is straightforward.  For each
particle $i$, let $\hat{\Vec{r}}_i$ and $\hat{\Vec{v}}_i$ be two
vectors drawn from a uniform distribution on the unit sphere
$\mathbf{S}^2$.  Then the position of particle $i$ is $\Vec{r}_i = r_i
\, \hat{\Vec{r}}_i$, where $r_i$ is chosen by drawing a random number
$x$ from a uniform distribution in the range $[0, m(\infty)]$ and
solving $m(r_i) = x$, and the velocity is $\Vec{v}_i = v_i \,
\hat{\Vec{v}}_i$, where $v_i$ is chosen from the speed distribution
$v^2 f(\frac{1}{2} v^2 + \Phi(r_i))$ using rejection sampling
\citep{vN51}.  The particle masses are $m_i = m(\infty) / N_{\rm
sphr}$.

The $N_{\rm test}$ test particles are initially placed in a single
disk.  If the test particle distribution follows $m_{\rm d}(r)$, the
initial radius $q_i$ of particle $i$ may be selected by drawing a
random number $x$ from a uniform distribution in the range $[0, m_{\rm
d}(\infty)]$ and solving $m_{\rm d}(q_i) = x$.  However, we prefer to
bias the distribution by a factor of $r^2$ to improve disk sampling at
large $r$ (see \S~\ref{sec:methodology}).  Let
\begin{equation}
  \eta(r) =
    \int_0^r d \chi \, \chi^2 \, m_{\rm d}^{\prime}(\chi) \, ,
\end{equation}
where $m_{\rm d}^{\prime}(r) = d m_{\rm d} / d r$; then $q_i$ is
selected by drawing $x$ from $[0,\eta(\infty)]$ and solving $\eta(q_i)
= x$.  The orbital velocity $v_i$ of particle $i$ is calculated using
the smoothed profile:
\begin{equation}
  v_i = \sqrt{G \, \overline{m}(q_i) / q_i} \, .
  \label{eq:sphr-circ-vel}
\end{equation}
Note that test particles are placed on {\it exactly\/} circular
orbits, creating a perfectly ``cold'' disk.  Finally, the position
$\Vec{r}_i$ and velocity $\Vec{v}_i$ of particle $i$ are rotated to
align $\Vec{r}_i \times \Vec{v}_i$ with a normalized angular momentum
$\hat{\Vec{s}}_i$ drawn from a uniform distribution on $\mathbf{S}^2$.

\section{Random Merger Models}

Our galaxy construction procedure has some elements in common with
\citet{MD07}.  Like them, we compute isotropic distribution functions
$f_{\rm b}(E)$ and $f_{\rm h}(E)$ for the bulge and halo,
respectively, by approximating the disk's gravitational field with its
spherically averaged equivalent.  Unlike them, we use the resulting
bulge and halo ``as is'', without first adiabatically imposing a
flattened disk potential; the response of the bulge and halo to such
an adiabatic transformation is so subtle that a good approximation to
equilibrium is possible without it.  This makes our procedure quite
fast.  What follows is a brief technical description of our procedure;
a full discussion and numerical tests will be presented elsewhere
(Barnes, in preparation).

The bulge follows a \citet{H90a} model out to a radius $b_{\rm b}$,
and tapers at larger radii to avoid placing a small number of
particles at extremely large distances:
\begin{equation}
  \rho_{\rm b}(r) =
    \left\{
      \begin{array}{ll}
	\displaystyle
          \frac{a_{\rm b} m_{\rm b}}{2 \pi} \,
	      \frac{1}{r (a_{\rm b} + r)^{3}} \, , &
	    r \le b_{\rm b} \\ [0.4cm]
	\displaystyle
	  \rho_{\rm b}^{*} \, \left(\frac{b_{\rm b}}{r}\right)^2 \,
	      e^{-2 r / b_{\rm b}} \, , &
	    r > b_{\rm b} \\
      \end{array}
    \right.
  \label{eq:bulge-model}
\end{equation}
where $m_{\rm b}$ is the bulge mass, and $\rho_{\rm b}^{*}$ is fixed
by requiring that $\rho_{\rm b}(r)$ be continuous at $r = b_{\rm b}$.
For $b_{\rm b} \gg a_{\rm b}$, the slope $d \rho_{\rm b} / d r$ is
also continuous at $r = b_{\rm b}$.

The halo follows a \citet{NFW96} model out to a radius $b_{\rm h}$,
and tapers at larger radii as proposed by \citet{SW99}:
\begin{equation}
  \rho_{\rm h}(r) =
    \left\{
      \begin{array}{ll}
	\displaystyle
	  \frac{m_{\rm h}(a_{\rm h})}{4 \pi (\ln(2) - \frac{1}{2})} \,
	      \frac{1}{r (r + a_{\rm h})^2} \, , &
	    r \le b_{\rm h} \\ [0.4cm]
	\displaystyle
	  \rho_{\rm h}^{*} \, \left(\frac{b_{\rm h}}{r}\right)^\beta \,
	      e^{- r / a_{\rm h}} \, , &
	    r > b_{\rm h} \\
      \end{array}
    \right.
  \label{eq:halo-model}
\end{equation}
where $m_{\rm h}(a_{\rm h})$ is the halo mass within radius $a_{\rm
h}$, and $\rho_{\rm h}^{*}$ and $\beta$ are fixed by requiring that
$\rho_{\rm h}(r)$ and $d \rho_{\rm h} / d r$ are both continuous at $r
= b_{\rm h}$.  The halo is tapered more abruptly than the bulge to
tame the logarithmic divergence of the standard \citeauthor{NFW96}
mass profile as $r \to \infty$.

The disk has an exponential radial profile \citep{dV59a, dV59b, F70}
and a $\mathrm{sech}^2$ vertical profile \citep{vdKS81}:
\begin{equation}
  \rho_{\rm d}(q,\phi,z) =
    \frac{m_{\rm d}}{4 \pi a_{\rm d}^2 z_{\rm d}} \,
      e^{- q / a_{\rm d}} \,
        \mathrm{sech}^2 \left( \frac{z}{z_{\rm d}} \right) \, ,
  \label{eq:disk-model}
\end{equation}
where $(q = \sqrt{x^2+y^2}, \phi, z)$ are cylindrical coordinates and
$m_{\rm d}$ is the total mass of the disk.

For the bulge and halo, cumulative mass profiles are obtained by
integrating (\ref{eq:bulge-model}) and (\ref{eq:halo-model}):
\begin{equation}
  m_{\rm b}(r) = \int_0^r d\chi \, 4 \pi \chi^2 \, \rho_{\rm b}(\chi) \, ,
  \qquad {\rm and} \qquad
  m_{\rm h}(r) = \int_0^r d\chi \, 4 \pi \chi^2 \, \rho_{\rm h}(\chi) \, .
\end{equation}
We use the cumulative mass profile for an infinitely thin disk,
\begin{equation}
  m_{\rm d}(r) =
    m_{\rm d} \, (1 - e^{- r / a_{\rm d}}) (1 + r / a_{\rm d}) \, ,
\end{equation}
which is adequate for our purposes since $m_{\rm b}(r) \gg m_{\rm
d}(r)$ at small $r$.  These functions are summed to get the total mass
profile $m(r)$, the smoothed profile $\overline{m}(r)$ is computed
using (\ref{eq:smooth-mass-profile}), and the potential $\Phi(r)$ is
computed using (\ref{eq:potential-equation}).  We then express
$\rho_{\rm b}$ and $\rho_{\rm h}$ as functions of $\Phi$, insert these
functions in (\ref{eq:eddington-formula}) to obtain $f_{\rm b}(E)$ and
$f_{\rm h}(E)$, and construct N-body realizations of the bulge and
halo following the procedure described in Appendix~A.

The disk is realized by sampling an approximate distribution function
\begin{equation}
  f_{\rm d}(q,\phi,z,v_{\rm q},v_\phi,v_{\rm z}) \, \propto \, 
    \rho_{\rm d}(q,\phi,z) \;
    \mathcal{H}\!\left[\frac{v_{\rm q}}{\sigma_{\rm q}(q)}\right] \,
    \mathcal{H}\!\left[\frac{v_\phi-\overline{v}(q)}{\sigma_\phi(q)}\right] \,
    \mathcal{G}\!\left[\frac{v_{\rm z}}{\sigma_{\rm z}(q)} \right] \, ,
  \label{eq:disk-dist-func}
\end{equation}
where $v_{\rm q}$, $v_\phi$ and $v_{\rm z}$ are velocities in the
radial, azimuthal, and vertical directions, respectively.  The
function $\overline{v}(q)$ is the mean rotation velocity, while
$\sigma_{\rm q}(q)$, $\sigma_\phi(q)$, and $\sigma_{\rm z}(q)$ are
dispersions in the radial, azimuthal, and vertical directions,
respectively.  The function $\mathcal{G}(x)$ is a Gaussian, while
$\mathcal{H}(x)$ resembles $\mathcal{G}(x)$ but cuts off faster for
large $|x|$:
\begin{equation}
  \mathcal{G}(x) \propto e^{- \frac{1}{2} x^2} \, ,
  \qquad \qquad
  \mathcal{H}(x) \propto e^{- \frac{1}{2} (x/c)^2 - \frac{1}{4} (x/c)^4} \, ,
\end{equation}
where $c$ is fixed by requiring $\int dx \, x^2 \, \mathcal{H}(x) =
\int dx \, \mathcal{H}(x)$.  This function is used instead of a
Gaussian to avoid overpopulating the high-velocity tail of the
distribution.

The local circular velocity $v_{\rm c}(q)$ is given by
\begin{equation}
  v_{\rm c}^2(q) =
    G \, \frac{\overline{m}_{\rm s}(q)}{q} +
      q \, \frac{d \Phi_{\rm d}}{d q} \, ,
  \label{eq:circular-vel}
\end{equation}
where $\overline{m}_{\rm s}(q)$ is the smoothed spheroid (bulge +
halo) mass profile and $\Phi_{\rm d}$ is the potential due to the
disk.  To compute $\overline{m}_{\rm s}(r)$ we insert the spheroid
profile $m_{\rm s}(r) = m_{\rm b}(r) + m_{\rm h}(r)$ in
(\ref{eq:smooth-mass-profile}).  Our expression for $\Phi_{\rm d}$
explicitly takes ``softening'' into account:
\begin{equation}
  \frac{d \Phi_{\rm d}}{d q} = - \, \frac{G  m_{\rm d}}{a_{\rm d}^3} \,
    \int_0^\infty dk \,
    \frac{k\,e^{-k\epsilon_{\rm d}}\,J_1(k q)}{(a_{\rm d}^{-2}+k^2)^{3/2}} \, ,
  \label{eq:smoothed-disk-accel}
\end{equation}
where $J_1(x)$ is the cylindrical Bessel function of order one, and
setting $\epsilon_{\rm d} = \sqrt{\epsilon^2 + z_{\rm d}^2}$ allows --
in an approximate way -- for the finite thickness of the disk.

The vertical dispersion is given by the solution for an isothermal
sheet \citep[e.g.,][p.~282]{BT87}:
\begin{equation}
  \sigma_{\rm z}(q) = \sqrt{\pi G z_{\rm d} \Sigma(q)} \, ,
  \label{eq:sigma-z}
\end{equation}
where $\Sigma(q) = \int dz \, \rho_{\rm d}(q,z)$ is the surface
density of the disk at cylindrical radius $q$.  The radial dispersion
is then determined by fixing the ratio $\sigma_{\rm q}/\sigma_{\rm
z}$:
\begin{equation}
  \sigma_{\rm q}(q) =
    \mu(q) \, \sigma_{\rm z}(q) =
      \left(1 + \frac{q}{q + q_\sigma}\right) \, \sigma_{\rm z}(q) \, ,
  \label{eq:sigma-q}
\end{equation}
where $q_\sigma$ is a scale parameter comparable to $a_{\rm d}$.
The function $\mu(q)$ is chosen to make $\sigma_{\rm q}/\sigma_{\rm z}
\simeq 2$ in the body of the disk -- roughly matching the solar
neighborhood value \citep[e.g.,][]{DB98} -- while letting $\sigma_{\rm
q}/\sigma_{\rm z} \to 1$ for $q \to 0$.  The azimuthal dispersion is
related to the radial dispersion \citep[][p.~203]{BT87}:
\begin{equation}
  \sigma_\phi(q) = \frac{\kappa(q)}{2 \Omega(q)} \, \sigma_{\rm q}(q) \, ,
  \label{eq:sigma-phi}
\end{equation}
where $\Omega(q) = v_{\rm c}(q) / q$ is the circular orbital frequency
and $\kappa(q) = \sqrt{4 \Omega^2 + q d \Omega^2 / d q}$ is the
epicyclic frequency.

Finally, the mean rotation velocity $\overline{v}(q)$ is determined
using the axisymmetric Jeans equation \citep[e.g.,][p.~198]{BT87}:
\begin{equation}
  \overline{v}^2(q) =
    v_{\rm c}^2(q) +
      \sigma_{\rm q}^2(q) \, \left(1 - \frac{2 q}{a_{\rm d}}\right) -
        \sigma_\phi^2(q) +
          \sigma_{\rm z}^2(q) \, q \, \frac{d \mu^2}{d q}
  \label{eq:mean-velocity}
\end{equation}

The parameter values needed to completely define the galaxy model are:
\begin{equation}
  \begin{array}{l@{\quad}l@{\quad}l}
    m_{\rm b} = 0.0625 \, , &
    a_{\rm b} = 0.02 \, , &
    b_{\rm b} = 4.0 \, , \\ 
    m_{\rm d} = 0.1875 \, , &
    a_{\rm d} = 1/12 \, , &
    z_{\rm d} = 0.0075 \, , \\
    m_{\rm h}(a_{\rm h}) = 0.16 \, , &
    a_{\rm h} = 0.25 \, , &
    b_{\rm h} = 0.98015 \, , \\
    \epsilon = 0.0075 \, , &
    \tilde{\epsilon} = 0.0115 \, , &
    \tilde{\epsilon}_{\rm s} = 0.0115 \, , \\
    q_\sigma = 0.075  \, , &
    \kappa = 1.975 \, , &
    \kappa_{\rm s} = 2.025 \, .
  \end{array}
\end{equation}
A few remarks about these parameters are in order.  First, tapering
the bulge as in (\ref{eq:bulge-model}) with $b_{\rm b} = 200 a_{\rm
b}$ reduces the total bulge mass by $\sim 0.5$\%; to correct this, the
value of $m_{\rm b}$ actually used in (\ref{eq:bulge-model}) is
adjusted upward accordingly.  Second, the primary halo mass parameter
is $m_{\rm h}(a_{\rm h})$; the halo taper radius $b_{\rm h}$ is
adjusted to make the total halo mass $m_{\rm h}(\infty) = m_{\rm h} =
1.0$.  Third, the softening parameter actually used in the N-body
calculations is $\epsilon$; the values of $\tilde{\epsilon}$ and
$\kappa$ listed here are chosen by computing $\Phi(r)$ using
(\ref{eq:potential-equation}) and comparing the result to an N-body
calculation.  Fourth, the parameters $\tilde{\epsilon}_{\rm s}$ and
$\kappa_{\rm s}$ used to compute the smoothed spheroid mass profile
$\overline{m}_{\rm s}(r)$ are likewise chosen by comparison with an
N-body calculation.

\begin{figure}[t!]
\begin{center}
\includegraphics[clip=true,width=0.5\columnwidth]{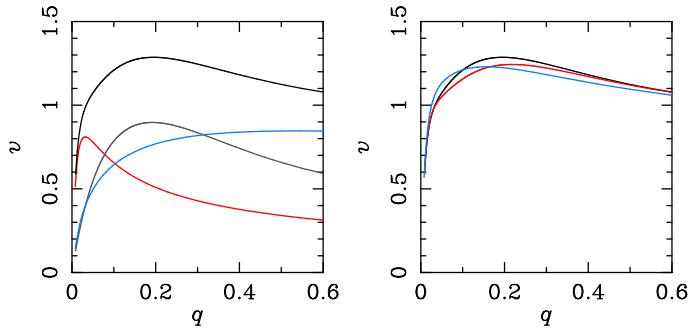}
\caption{Circular velocity profiles for the disk galaxy model.  Left:
circular velocities for the bulge (red), disk (grey), and halo (blue).
The total circular velocity is also shown (black).  Right: total
circular velocity (black), mean rotation velocity (red), and circular
velocity for the spherical mass model (blue).
\label{galmod}}
\end{center}
\end{figure}

Fig.~\ref{galmod} presents circular velocity profiles for the galaxy
model adopted here.  The left-hand panel shows profiles for the
individual components, computed taking softening into account as
described above.  Also shown is the total circular velocity $v_{\rm
c}(q)$ given by (\ref{eq:circular-vel}).  The right-hand panel again
shows $v_{\rm c}(q)$ and compares it with the mean rotation velocity
$\overline{v}(q)$ given by (\ref{eq:mean-velocity}) and the circular
velocity for the equivalent spherical mass model given by
(\ref{eq:sphr-circ-vel}).

To check this model, we constructed a realization with $N_{\rm b} =
16384$ bulge particles, $N_{\rm d} = 49152$ disk particles, and
$N_{\rm h} = 65536$ halo particles.  This system was then evolved in
isolation for $10$ time units, using a hierarchical N-body
code\footnote{See
\url{http://www.ifa.hawaii.edu/faculty/barnes/treecode/treeguide.html}
for a discussion of this code, which generalizes earlier modifications
\citep{B90} of the original tree code \citep{BH86}.} with an accuracy
parameter $\theta = 1$, quadrupole-moment corrections \citep{H87}, a
Plummer softening length $\epsilon = 0.0075$, and a leap-frog
integrator with a time-step $\Delta t = 1/256$.  During the first
$0.25$ time units the ratio of kinetic to potential energy, $T/U$,
fell from an initial value of $0.4980$ to $0.4945$; it then fluctuated
around this value with an amplitude of $\sim 0.003$.  This initial
drop indicates that the model was not started in perfect equilibrium,
but the implied rearrangement of mass is only a little larger than the
$1/\sqrt{N}$ ($\simeq 0.0028$) fluctuations occurring in an N-body
system with this $N$.  Apart from transient spiral structure, this
model showed no significant features until the disk begins to develop
a bar at time $t \simeq 4$.  It's hard to completely suppress a weak
bar instability in galaxy models with relatively massive disks like
the one used here; however, this instability has little effect on the
merger simulations since the galaxies interact with each other long
before they would develop bars in isolation.

The $36$ random merger simulations were run with the same number of
particles per galaxy and N-body integration parameters used in the
test just described.  Energy was conserved to $\sim 0.05$\% even in
the most violent encounters.  Particle positions and velocities were
output every $\Delta t_{\rm out} = 1/32$ time units, providing a large
data base which could be used to construct random samples like the one
in Fig.~\ref{random_mergers}.

\section{Velocity Scale Bias}

As noted at the end of \S~\ref{sec:parameters}, Identikit estimates of
the velocity scale $\mathcal{V}$ are typically $\sim 10$\% too high.
The value of $\mathcal{V}$ is usually determined toward the end of the
matching process; after values have been selected for most other
parameters, we adjust $\mathcal{V}$ to obtain a good overall match
between the particles and the grey-scale images in the $(X,V)$ and
$(V,Y)$ planes.  There are a number of factors which may influence the
choice for $\mathcal{V}$, including the velocity difference between
the galaxies and the characteristic velocities of tidal features.
However, a key feature is the velocity widths of the galaxies; we
generally try to adjust $\mathcal{V}$ so that the particles span the
full range of velocities present in each galaxy.

\begin{figure}[t!]
\begin{center}
\includegraphics[clip=true,width=0.45\columnwidth]{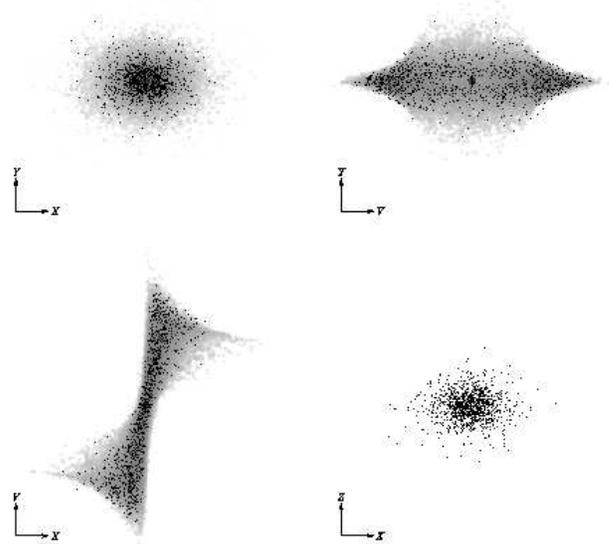}
\caption{Identikit match to a single disk galaxy.  The velocity scale
  factor $\mathcal{V}_{\rm fit}$ was adjusted by eye to match the
  width of the velocity distribution.
  \label{fitvscale}}
\end{center}
\end{figure}

While matching the velocity widths of the individual galaxies {\it
seems\/} reasonable, it appears to be the source of the bias in
$\mathcal{V}$.  Velocities in the test-particle Identikit disks are
not identical to those in the self-gravitating disks used in the
random mergers; the former are perfectly ``cold'', while the latter
have non-zero velocity dispersions.  In addition, as the right-hand
panel in Fig.~\ref{galmod} shows, the circular velocity profile of the
Identikit model rises more rapidly, peaks at a smaller radius, and
then falls slightly below either the circular ($v_{\rm c}(q)$) or mean
($\overline{v}(q)$) profiles in the self-gravitating model.

Fig.~\ref{fitvscale} shows an Identikit fit to a self-consistent disk
galaxy model.  This disk is tilted by $45^\circ$ to the line of sight;
in fitting this model, the orientation and length scale factor were
first set to their actual values.  We then adjusted $\mathcal{V}_{\rm
fit}$ by eye, stretching the particle distribution to match the
apparent velocity width of the grey-scale images in the $(X,V)$ and
$(V,Y)$ projections.  The value of $\mathcal{V}_{\rm fit}$ selected in
this manner is $10.9$\% too high, much as in the Identikit matches to
interacting galaxies.  In hindsight, we note that the $(X,V)$
projection shows a handful of points falling outside the rotation
curve for large $|X|$; a fit giving priority to these points would
have recovered a more accurate value for $\mathcal{V}$.

Compared to its Identikit analog, the larger velocity width of the
self-consistent disk is a direct consequence of the random velocities
imparted to individual disk particles.  To test this, we constructed
self-consistent disks with different dispersions and velocity scales,
and selected best-fit values of $\mathcal{V}_{\rm fit}$ as in
Fig.~\ref{fitvscale}.  The ratio $\mathcal{V}_{\rm fit} /
\mathcal{V}_{\rm true}$ has a one-to-one relationship with the disk's
velocity dispersion; the larger the dispersion, the greater the
overestimate of $\mathcal{V}$.  It's likely that by constructing
Identikit models with random velocities comparable to those present in
the self-gravitating disks we could largely remove this bias.
However, ``cold'' disks may actually be more appropriate when fitting
kinematics observed in cold gas tracers (e.g., H{\small I} or CO),
since these have smaller velocity dispersions than most stellar
components.


\clearpage

\begin{deluxetable}{lcrrr}
\tablewidth{0pt}
\tablecaption{Ratios of fit/true parameter values for all $36$ models.
              \label{fit_ratios}}
\tablehead{\colhead{{\bf Parameter}}&
	   \colhead{}&
	   \colhead{{\bf 1$^{\rm st}$ Quart}}&
           \colhead{{\bf Median}}&
	   \colhead{{\bf 3$^{\rm rd}$ Quart}}}
\startdata
time since pericenter  &      $t$      & $0.86$ & $1.02$ & $1.14$ \\
pericentric separation &      $p$      & $0.79$ & $1.01$ & $1.25$ \\
length scale factor    & $\mathscr{L}$ & $0.86$ & $1.01$ & $1.17$ \\
velocity scale factor  & $\mathcal{V}$ & $1.05$ & $1.10$ & $1.15$ \\ [0.25cm]
physical time          &      $T$      & $0.78$ & $0.95$ & $1.02$ \\
physical separation    &      $P$      & $0.87$ & $0.97$ & $1.13$ \\
\enddata
\end{deluxetable}

\begin{deluxetable}{lrrccl}
\tabletypesize{\scriptsize}
\tablecaption{Dynamical models of interacting disk galaxies\tablenotemark{a}.
              \label{dynamical_models}}
\tablewidth{0pt}
\tablehead{\colhead{{\bf System}\tablenotemark{b}}&
	   \colhead{{\bf Arp}}&
	   \colhead{{\bf VV}}&
	   \colhead{{\bf Type}}&
           \colhead{{\bf Match}\tablenotemark{c}}&
	   \colhead{{\bf Reference}}}
\startdata
VV 784                 &     & 784 & ring   & morph & \citet{T78} \\
                       &     &     &        & kin1  & \citet{SMH93} \\
VV 347, Arp 119        & 119 & 347 & ring   & qual  & \citet{HL01} \\
NGC 520                & 157 & 231 & S+S    & kin1  & \citet{SB91} \\
3C 48                  &     &     & S+S    & kin1  & \citet{SEPZKS04} \\
NGC 672 / IC 1727      &     & 338 & S+S    & kin2  & \citet{CFWG80}  \\
NGC 1143/44            & 118 & 331 & E+S    & qual  & \citet{LHG88} \\
IC 1908, AM 0313-545   &     &     & S+d    & qual  & \citet{MBR93} \\
NGC 2207 / IC~2163     &     &     & S+S    & qual  & \citet{ESKBE95} \\
                       &     &     &        & kin2  & \citet{SKBTEE05} \\
VV 785, AM 0644-741    &     & 785 & ring   & kin1  & \citet{AW07} \\
NGC 2442 / AM 0738-692 &     &     & S+d    & qual  & \citet{MB97} \\
II Hz 4                &     &     & ring   & qual  & \citet{LT76} \\
NGC 2782               & 215 &     & S+?\tablenotemark{d}
                                                    & qual  & \citet{S94} \\
NGC 2992/93            & 245 &     & S+S    & kin2  & \citet{DBSPWM00} \\
NGC 3031/77            &     &     & S+d    & qual  & \citet{TD93} \\
NGC 3031/34/77         &     &     & S+S+d  & qual  & \citet{BBCKB91} \\
                       &     &     &        & qual  & \citet{Y99} \\
AM 1003-435            &     &     & S+S    & qual  & \citet{GARD06} \\
NGC 3395/96            & 270 & 246 & S+S    & kin1  & \citet{CBAG99} \\
NGC 3448 / UGC 6016    & 205 &     & S+d    & qual  & \citet{NK86} \\
NGC 4038/39            & 244 & 245 & S+S    & morph & \citet{TT72} \\
                       &     &     &        & kin2  & \citet{vdH79} \\
                       &     &     &        & qual  & \citet{MvdHB87} \\
                       &     &     &        & qual  & \citet{B88} \\
                       &     &     &        & qual  & \citet{MBR93} \\
                       &     &     &        & kin2  & \citet{H03} \\
NGC 4254/92?\tablenotemark{e}
                       &     &     & S+S    & qual  & \citet{DB08} \\
NGC 4435/38            & 120 & 188 & S0+S   & morph & \citet{CDCP88} \\
                       &     &     &        & kin2  & \citet{VBCS05} \\
NGC 4449 / DDO 125     &     &     & S+d    & gen   & \citet{TK01} \\
NGC 4654/39?\tablenotemark{e}
                       &     &     & S+S    & kin2  & \citet{V03} \\
NGC 4676               & 242 & 224 & S+S    & morph & \citet{TT72} \\
                       &     &     &        & qual  & \citet{S74} \\
                       &     &     &        & kin1  & \citet{MBR93} \\
                       &     &     &        & qual  & \citet{GS94} \\
                       &     &     &        & kin1  & \citet{SR98} \\
                       &     &     &        & kin2  & \citet{B04} \\
NGC 5194/95            & 085 & 001 & S+S0   & morph & \citet{TT72} \\
                       &     &     &        & kin2  & \citet{T78} \\
                       &     &     &        & qual  & \citet{H90b}, \\
                       &     &     &        & kin2  & \citet{SL00} \\
                       &     &     &        & gen   & \citet{WD01} \\
                       &     &     &        & gen   & \citet{TS03} \\
                       &     &     &        & kin2  & \citet{DMFJC03} \\
NGC 5216/18            & 104 & 033 & E+S    & qual  & \citet{CAGS07} \\
NGC 5394/95            & 084 & 048 & S+S    & qual  & \citet{KBEEKST99}  \\
AM 2004-662            &     &     & E+d    & kin1  & \citet{DRDC00} \\
NGC 6872 / IC~4970     &     &     & S+S0   & qual  & \citet{MBR93} \\
                       &     &     &        & qual  & \citet{HK07} \\
NGC 7252               & 226 &     & S+S    & qual  & \citet{BR91} \\
                       &     &     &        & qual  & \citet{MBR93} \\
                       &     &     &        & kin2  & \citet{HM95} \\
                       &     &     &        & qual  & \citet{MDH98} \\
NGC 7714/15            & 284 & 051 & ring   & qual  & \citet{SW92} \\
                       &     &     &        & kin2  & \citet{SS03} \\
NGC 7752/53            & 086 & 005 & S+d    & kin2  & \citet{SL93} \\
\enddata

\tablenotetext{a}{This table attempts to survey and characterize dynamical
                  models of interacting galaxies which make
                  significant use of kinematic constraints.  Different
                  authors often use very different criteria when
                  imposing kinematic constraints, and published
                  descriptions are sometimes ambiguous.  No warranty
                  of completeness is expressed or implied.}

\tablenotetext{b}{In this column, a slash separates components of a
	          given system, while a comma separates alternate
	          names.  NGC numbers are used when available; if all
	          components have NGC numbers, the full number is
	          given for the first galaxy, and only the last two
	          digits for the rest.}

\tablenotetext{c}{Briefly, ``kin2'' matches are constrained by 2-D
                  kinematic data, ``kin1'' matches are constrained by
                  1-D kinematic data, ``gen'' matches used genetic
                  algorithms, ``qual'' matches reproduce qualitative
                  kinematic features, and ``morph'' matches reproduce
                  morphology (listed {\it only\/} as precursors).  See
                  text for details.}

\tablenotetext{d}{Type of companion is ambiguous.}

\tablenotetext{e}{Identity of companion is ambiguous.}

\end{deluxetable}

\end{document}